\documentclass[10pt]{article}

\usepackage{amsmath,amssymb,amsthm}
\usepackage{graphicx}
\usepackage{microtype}
\usepackage{rotating}
\usepackage{array}
\usepackage{graphicx}
\usepackage{epsfig}
\usepackage{algorithmic}
\usepackage{placeins}
\usepackage{marginnote}

\usepackage{./pca}
\usepackage{./IEEEeqnarray}
\usepackage{./progSTF}
\usepackage{./polyhedraSTF,./matrixSTF}

\usepackage{syntax}

\theoremstyle{definition}

\newcommand{\edt}{{\small EDT}}
\newcommand{\edts}{{\small EDT}s}
\newcommand{\ocr}{{\small OCR}}

\newcommand{\tbb}{{\small TBB}}
\newcommand{\ir}{{\small IR}}

\newcommand{\quast}{{\small QUAST}}

\newcommand{\swarm}{{\small SWARM}}

\newcommand{\cnc}{{\small C}n{\small C}}
\newcommand{\DAG}{{\small DAG}}
\newcommand{\ral}{{\small RAL}}
\newcommand{\omp}{{\small O}pen{\small MP}}
\newcommand{\stapl}{{\small STAPL}}

\newcommand{\worker}{{\small WORKER}}
\newcommand{\prescriber}{{\small PRESCRIBER}}

\newcommand{\Startup}{{\small STARTUP}}
\newcommand{\Worker}{{\small WORKER}}

\newcommand{\Shutdown}{{\small SHUTDOWN}}
\newcommand{\TaskTag}{{\small TASKTAG}}
\newcommand{\cloog}{{\small CLOOG}}

\newcommand{\src}{{\sf s}}
\newcommand{\dst}{{\sf t}}

\newcommand{\divergencethreedoneiteration}{{\small DIV}-3D-1}%{\small IT}}
\newcommand{\fdtdtwod}{{\small FDTD}-2D}
\newcommand{\gaussseideltwodfivepointsenlarged}{{\small GS}-2D-5P}
\newcommand{\gaussseideltwodninepointsenlarged}{{\small GS}-2D-9P}
\newcommand{\gaussseidelthreedtwosevenpointsenlarged}{{\small GS}-3D-27P}
\newcommand{\gaussseidelthreedsevenpointsenlarged}{{\small GS}-3D-7P}
\newcommand{\jactwod}{{\small JAC}-2D-{\small COPY}}
\newcommand{\jacobidoubletwodfivepointsenlarged}{{\small JAC}-2D-5P}
\newcommand{\jacobidoubletwodninepointsenlarged}{{\small JAC}-2D-9P}
\newcommand{\jacobidoublethreedtwosevenpointsenlarged}{{\small JAC}-3D-27P}
\newcommand{\jacobidoublethreedsevenpointsenlargedoneiteration}{{\small JAC}-3D-1}%{\small IT}}
\newcommand{\jacobidoublethreedsevenpointsenlarged}{{\small JAC}-3D-7P}
\newcommand{\lud}{{\small LUD}}
\newcommand{\matmult}{{\small MATMULT}}
\newcommand{\pmatmult}{{\small P-MATMULT}}
\newcommand{\poisson}{{\small POISSON}}
\newcommand{\rtmthreed}{{\small RTM}-3D}

\newcommand{\sor}{{\small SOR}}
\newcommand{\strsm}{{\small STRSM}}
\newcommand{\trisolv}{{\small TRISOLV}}

\newcommand{\cncdepend}{{\small DEP}}
\newcommand{\cncsyncget}{{\small BLOCK}}
\newcommand{\cncunsafeget}{{\small ASYNC}}
\newcommand{\openmp}{{\small OMP}}

\definecolor{keywordcolor}{rgb}{0.4,0,0}
\renewcommand{\KEYWORD}[1]{{\color{keywordcolor}\bf #1}\dontgobblespace}
\newcommand{\IF}{\KEYWORD{if}}
\newcommand{\AND}{\KEYWORD{and}}
\newcommand{\OR}{\KEYWORD{or}}
\newcommand{\THEN}{\KEYWORD{then}}
\newcommand{\ELSE}{\KEYWORD{else}}

\newcommand{\ENDIF}{\KEYWORD{end if}}

\newcommand{\REPEAT}{\KEYWORD{repeat}}
\newcommand{\UNTIL}{\KEYWORD{until}}

\begin{document}

\setlength{\pdfpageheight}{\paperheight}
\setlength{\pdfpagewidth}{\paperwidth}

{\centering

{\bfseries\Large A Tale of Three Runtimes\bigskip}

Nicolas Vasilache, Muthu Baskaran, Tom Henretty, Benoit Meister,\\
M. Harper Langston, Sanket Tavarageri, Richard Lethin \\
Reservoir Labs Inc. 632 Broadway, New York, NY, 10012\footnote{
This material is based upon work supported by the Department of Energy
under Award Number(s) No. DE-SC0008717.

This report was prepared as an account of work sponsored by an agency of
the United States Government. Neither the United States Government nor
any agency thereof, nor any of their employees, makes any warranty,
express or implied, or assumes any legal liability or responsibility for
the accuracy, completeness, or usefulness of any information, apparatus,
product, or process disclosed, or represents that its use would not
infringe privately owned rights. Reference herein to any specific
commercial product, process, or service by trade name, trademark,
manufacturer, or otherwise does not necessarily constitute or imply its
endorsement, recommendation, or favoring by the United States Government
or any agency thereof. The views and opinions of authors expressed
herein do not necessarily state or reflect those of the United States
Government or any agency thereof.

This report describes Patent Pending Technology.} \\
January 30, 2013\\
}
%           {vasilache@reservoir.com, baskaran@reservoir.com, henretty@reservoir.com, meister@reservoir.com, langston@reservoir.com, lethin@reservoir.com}
%\authorinfo{Authorship omitted for blind review purposes}
%           {Institution}
%           {xxx@yyy.com}

% \maketitle

\begin{abstract}
This contribution discusses the automatic generation of event-driven,
tuple-space based programs for task-oriented execution models from a sequential C specification. 
We developed a hierarchical mapping solution using 
auto-parallelizing compiler technology to target three different runtimes
relying on event-driven tasks (\edts). 
Our solution benefits from the important observation that loop types encode
short, transitive relations among \edts\ that are compact and efficiently
evaluated at runtime. In this context, permutable loops are of particular
importance as they translate immediately into conservative point-to-point
synchronizations of distance $1$. 
Our solution generates calls into a runtime-agnostic C++ layer, which we
have retargeted to Intel's Concurrent Collections (\cnc), ETI's \swarm,
and the Open Community Runtime (\ocr). 
Experience with other runtime systems motivates our introduction of support
for hierarchical async-finishes in \cnc. 
Experimental data is provided to show the benefit of automatically
generated code for \edt-based runtimes as well as comparisons across runtimes.
%to compare our work with OpenMP.
%Comparisons with \omp\ codes auto-parallelized by Reservoir Labs'
%\RStream\ compiler are provided.
\end{abstract}

% general terms are not compulsory anymore, you may leave them out

% \category{D.3.4}{Programming Languages}{Processors -- Code Generation, 
%   Compilers, Optimization}
% \terms{Algorithms, Design, Performance}
% \keywords{Parallelization, Compiler Optimization, Polyhedral Model, EDT, codelet,
%    Work-Stealing, C++11, Expression Template, Automatic Translation}

\section{Introduction}
%Despite predictions on the end of Moore years, for both physical
%and economical reasons Intel has recently declared Moore's law alive and well. 
%However, as the number of transistors fitting a given chip area continues to
%grow, so does the energy required to enable them, resulting in the
%heat envelope supported by the packaging being reached. 
%The era of sequential computing relying on ever increasing clock speeds and
%decomposition of the processing pipeline into ever shorter stages has come
%to an end.
%As \gflops\ per Watt replaced traditional GHz,
%clock speeds stopped increasing and performance metrics started shifting.
%Subsequently, due to the same power wall which halted frequency
%scaling, the end of multi-core scaling was predicted.  
%Esmaeilzadeh et al. estimate that for any chip organization and
%topology, multi-core scaling will also be power limited~\cite{Esmaeilzadeh:2011}.
%To meet the power budget, they project ever more significant portions of
%the chip will have to be turned off to accommodate the increase in static power
%loss from increasing transistor count. We are entering the ``dark silicon'' era.

Hardware scaling considerations associated with the quest for
exascale and extreme scale computing are
driving system designers to consider event-driven-task (\edt)-oriented execution
models for executing on deep hardware hierarchies.  This 
paper describes a method for the automatic generation of optimized event-driven,
tuple-space-based programs from a sequential C specification. 
The work builds on the Reservoir Labs R-Stream compiler~\cite{DBLP:reference/parallel/MeisterVWBLL11,rstream30-Final}
and is implemented as an experimental feature in the R-Stream compiler.

We have developed a hierarchical mapping solution using auto-parallelizing
compiler technology to target three different runtimes 
based on \edts. Our solution consists of (1) a mapping strategy with selective trade-offs between parallelism and
  locality to extract fine-grained \edts, and (2)
 a retargetable runtime API that captures common aspects of the \edt\ programming
  model and uniformizes translation, porting, and comparisons between runtimes.
At a high level, complex loop nest restructuring transformations are applied
to construct a logical tree representation of the program. This representation
is mapped to a tree of \edts. 
Each \edt\  is associated with a unique (id, tag tuple) pair in the generated program.
Our solution generates calls into a runtime-agnostic C++ layer (\ral), which we
successfully retargeted to Intel's \cnc, ETI's \swarm, and the Open Community
Runtime (\ocr). We provide different experiments to characterize the
performance of our auto-generated code and %features of the 
targeted runtimes.

\hspace{-10mm}
\section{A Motivating Example}
\label{sec:motivation}

\begin{figure*}[htb] 
\centering

\begin{minipage}{\textwidth}
    \begin{rulett}
      \scriptsize{
for (t1=-1; t1<=\(\lfloor\frac{pT-2}{8}\rfloor\); t1++) \{ 
 lbp=max(\(\lceil\frac{t1}{2}\rceil\),\(\lceil\frac{16*t1-pT+3}{16}\rceil\)); ubp=min(\(\lfloor\frac{pT+pN-2}{16}\rfloor\),\(\lfloor\frac{8*t1+pN+7}{16}\rfloor\));
 doall (t2=lbp; t2<=ubp; t2++) \{         {\bf// par.}
  for (t3=max\((0,\lceil\frac{t1-1}{2}\rceil,\lceil\frac{16*t2-pN-14}{16}\rceil)\); 
   t3<=min\((\lfloor\frac{pT+pN-2}{16}\rfloor,\lfloor\frac{8*t1+pN+15}{16}\rfloor,\lfloor\frac{16*t2+pN+14}{16}\rfloor)\); t3++) \{
   // for t4, t5, t6, t7, t8 loops omitted
   if (t5 \% 2 == 0) S1(t5, t6, t7, t8);
   if (t5 \% 2 == 1) S2(t5, t6, t7, t8);
\}\}\}
        (a) \omp parallel version (8x16x16x128)}
    \end{rulett}
  \end{minipage}

\vspace*{0.5cm}

  \begin{minipage}{\textwidth}
    \begin{rulett}
      \scriptsize{
  for (t1=\(\lceil\frac{-pN-15}{16}\rceil\);t1<=\(\lfloor\frac{pT-3}{16}\rfloor\);t1++)\{ {\bf//perm}
    for (t2=max(t1,-t1-1);
     t2<=min\((\lfloor\frac{-8*t1+pT-2}{8}\rfloor,\lfloor\frac{8*t1+pN+7}{8}\rfloor,\lfloor\frac{pT+pN-2}{16}\rfloor)\);t2++)\{ {\bf//perm}
      for (t3=max\((0,\lceil\frac{t1+t2-1}{2}\rceil,\lceil\frac{16*t2-pN-14}{16}\rceil)\);
       t3<=min\((\lfloor\frac{pT+pN-2}{16}\rfloor,\lfloor\frac{16*t2+pN+14}{16}\rfloor,\lfloor\frac{8*t1+8*t2+pN+15}{16}\rfloor)\);t3++)\{ {\bf//perm}
       // for t4, t5, t6, t7, t8 loops omitted
       if (t5 \% 2 == 0) S1(t5, t6, t7, t8);
       if (t5 \% 2 == 1) S2(t5, t6, t7, t8);
  \}\}\}
        (b) \edt-ready permutable version (8x16x16x128)}
    \end{rulett}
  \end{minipage}

\caption{\label{fig:heat-3-d-codes}Diamond-tiled Heat-3D Kernel (parallel and permutable versions)}
%\end{figure}
%\}\}\}
%\begin{figure}[htb] 
%\caption{\label{fig:heat-3-d}Diamond-tiled Heat-3D Kernel by Bondhugula
%  et al.~\cite{BandishtiPB12}}
\end{figure*}

We begin by presenting a motivating example based on state-of-the-art
techniques applied to the heat equation kernel.
Previous contributions have highlighted the inherent scalability 
limitations of implicit stencils with time-tiling schemes~\cite{Sriram2007}.
Intuitively, the pipelined start-up cost needed to activate enough tasks to
feed $N$ processors grows linearly with the number of processors.
When $N$ is large, this translates into a sequential bottleneck.
On large machines, the traditional solution is to perform domain
decomposition and to exchange ghost regions in a near-neighbor collective
communication step.
A viable alternative for both scalable parallelism and locality is to use
diamond tiling on explicit stencils~\cite{BandishtiPB12}.
%or on implicit  stencils coupled with a red-black scheme. 
Still, even in the absence of obvious ill-balanced wavefront synchronizations,
significant load-balancing inefficiencies may remain.
For the purpose of illustration, we begin with the pseudo-code in
Figure~\ref{fig:heat-3-d-codes}; this code represents the outermost loops of a
diamond-tiled, 3D heat equation solver using an explicit 3D Jacobi
time step.
In this version, \verb|S1| reads an array \verb|A| and writes an array
\verb|B|, whereas \verb|S2| conversely reads array \verb|B| and writes array
\verb|A|\@. Experiments on the code in Figure~\ref{fig:heat-3-d-codes}(a) have
been recently published and demonstrate significant advantages over a domain
specific language~\cite{Tang:2011}.
The version in Figure~\ref{fig:heat-3-d-codes}(b) corresponds to the same code without
the ``tile-skewing'' transformation.\footnote{Time-skewing creates the \omp~parallel loop.
Since diamond tiling is used, the loop is well-balanced across
  multiple $t1$ iterations. }
As a consequence, the loop types of the inter-task loops are all
``permutable'' and contain only forward dependences~\cite{griebl04habilitation}.
We then apply the distributed dependence evaluation portion of our framework
described in section~\ref{sec:ral}.
To this end, we manually insert the above code into a \cnc\ skeleton 
and perform experiments on a 2-socket, 6-core-per-socket Intel Xeon 
E5-2620 running at 2.00~GHz. This port required minimal time once 
the infrastructure was established.
\begin{figure}[htb]
\centering
{\footnotesize
\begin{tabular}{l|lllllll}
\hline
Version / Procs & 1  & 2 & 3 & 4 & 6 & 8 & 12 \\
\hline
\hline
\omp & 14.90 & 8.01 & 6.09 & 4.44 & 3.36 & {\bf 2.86} & 3.16 \\
\cnc & 13.71 & 7.03 & 4.82 & 3.73 & 2.75 & {\bf 1.98} & 2.16 \\
\omp-N & 14.83 & 7.96 & 5.87 & 4.18 & 3.05 & 2.40 & {\bf 2.31}\\
\cnc-N & 13.79 & 6.85 & 4.59 & 3.48 & 2.44 & 1.89 & {\bf 1.42} 
\end{tabular}
}
\caption{Diamond tiling~\cite{BandishtiPB12} \omp~vs \cnc}
\end{figure}
We can draw a few conclusions from these first performance numbers.
First, there is a non-negligible benefit in single-thread mode
due to reduced complexity of the loop control flow (we avoid the
loop-skewing transformation which has impact down to the innermost loop).
Improved single-thread performance of the
\ral\ over \omp\ is a positive side-effect of our solution. Still,
  there is single-thread cost to parallelism: the base sequential
  runtime of the tiled code in figure~\ref{fig:heat-3-d-codes}(b) is 12.74s.
Second, even if diamond-tiled code does not suffer pipeline parallel overhead,
there are still benefits to be gained from load-balancing effects.
Third, placement of computations and data is crucial and will be an important
focus of future research. 
In this experiment, we pinned threads to cores and performed round-robin
pinning of pages to sockets with the \texttt{libnuma} library. Rows \omp-N and \cnc-N 
demonstrate the effect of NUMA pinning with still an approximate 40\% socket 
miss rate. Pinning remains uncontrolled in rows \omp\ and \cnc.
More controlled data placements are expected to produce better results and will be
needed as we progress to many sockets and longer off-core latencies.

\section{Position of the Problem}

As projected by~\cite{Kogge:Exascale08}, power limits in
the machine room are one of the principal constraints in reaching
exascale.  The strategy of relying solely on commodity server chips
and accelerators, as is the case for petascale computing, will result
in infeasible levels of power.  Consequently, new processor
and system architectures are being investigated and designed for exascale.
These new architectures will have new programming and execution models.

\edt\ execution models 
reincarnate~\cite{DBLP:journals/cacm/MyerS68} old ideas of dataflow~\cite{DBLP:conf/programm/Dennis74}
and macro-dataflow~\cite{DBLP:conf/isca/SpertusGSECD93}.  What is driving
this reincarnation of these old dataflow ideas is that the new exascale
hardware is expected to be powered with supply voltages that are near
the transistor threshold voltage~\cite{DBLP:conf/dac/KaulAHAKB12}.
Lowering supply voltage produces a quadratic improvement in power
efficiency of computing devices with only a linear slowdown in
throughput.  Consequently it is possible to get improved power
utilization {\it as long as an increase in parallelism can be found to
  offset the linear slowdown}.  Of course, pure dataflow has
the advantage of expressing the absolute maximum amount of
parallelism to the level of individual operations, but
that comes with too much overhead and does not solve
the scheduling problem~\cite{DBLP:conf/isca/CullerA88}.

Macro-dataflow groups individual operations into larger tiles and
provides scheduling algorithms with provably optimal
properties~\cite{Blumofe96}, with the possibility of an increase in
parallelism by virtue of it more accurately representing the minimal
set of synchronizations, compared to the current dominant execution
models utilizing bulk synchronization.  

Another important consequence
of lowering the supply voltage near threshold is that variations in
device performance are exacerbated.  Thus, beyond any intrinsic
imbalance from the application itself, the hardware will be
create imbalance.  This means that mechanisms for dynamic load balancing
increase in importance for near threshold computing (NTC).  

To address this, exascale architects envision that
additional degrees of parallelism must be found to overprovision the
execution units to aid in load
balancing~\cite{DBLP:conf/ipps/IancuHBZ10} and to hide remote access
latencies. This overprovisioning effect is already mainstream in the
GPGPU world, and in particular in CUDA, where a user specifies more
parallelism than can be exploited at any given time for the purpose of
hiding latencies.  In the CUDA programming model, this
overprovisioning still relies on the bulk-synchronous model at the
device level but has begun shifting to a task-based model at the
CPU-to-device level.

Another aspect of envisioned exascale architectures is that the
hierarchy of the hardware, which is currently at 3-5 levels (core,
socket, chassis, rack...) will extend to 10 or more levels (3-4 levels on
chip, and another 6+ levels of system packaging). In current
systems, execution models address these levels through loop blocking
or tiling~\cite{DBLP:conf/asplos/LamRW91}, and stacked heterogeneous execution
models (e.g., MPI+OpenMP).  Such approaches will become
cumbersome (program size, etc.) with the envisioned 10+ levels.  More
compact representations that are naturally
recursive~\cite{DBLP:conf/pldi/YiAK00} or cache
oblivious~\cite{DBLP:conf/focs/FrigoLPR99} are potentially better.

Some of these tradeoffs are described in one paper describing
the Intel Runnemede exascale research 
architecture~\cite{Carter2013Short}; the \ocr\ \cite{OCR} is designed
with Runnemede as one of the hardware targets it should run well on.

Traditional approaches to parallelism require the programmer to express 
it explicitly in the form of communicating sequential processes. The
fork-join model~\cite{openmp} and the bulk-synchronous
model~\cite{MPI-user-guide} are well-established methodologies for shared
and distributed memory systems respectively.  

The \edt\ model supports the combination of different styles of parallelism (data,
task, pipeline).
At a very high-level, the \edt\ program expresses computation tasks which can:
(1)produce and consume data,
(2) produce and consume control events, 
(3) wait for data and events, 
and (4) produce or cancel other tasks.
Dependences between tasks must be declared to the runtime, which
keeps distributed queues of ready tasks (i.e., whose dependences have all been
met) and decides where and when to schedule tasks. Work-stealing is used for
load-balancing purposes~\cite{Blumofe96}.
Specifying tasks and dependences satisfied
at runtime is common to \cnc ~\cite{CnC}, \ocr, \swarm\ \cite{swarm} and other \edt\ runtimes.
Unfortunately, few programs using these constructs are available to the
community.  Also, it is impractical to expect programmers to write
directly in \edt\ form; the expression of explicit dependences is
cumbersome, requiring a significant expansion in the number of 
lines of code, and opaque to visual inspection and debugging.
Direct \edt\ programming might be done in limited circumstances by some hero or uber
programmers, or in evaluation and experimentation with this execution model.

We describe a solution to automate the synthesis of \edt\ codes from simple
sequential codes.
%To this end, we use advanced code analysis and
%transformation strategies.

A viable transformation system for this purpose is based on the polyhedral
model. The transformation system proposed by~\cite{GV06} allows
very intricate transformation compositions, but its applicability is typically
limited by (among others) static dependence analysis. 
%In order to extract provably maximal parallelism, exact array dataflow is
%usually needed. Even though exact array dataflow analysis can support affine
%data-driven predicates through a tradeoff between memory footprint and 
%available parallelism, this is not enough to obtain a good coverage 
%of applications.
Solutions have been devised for expanding the scope of analyzable 
codes by (1) computing inter-procedural over- and 
under-approximations~\cite{Beatrice95interprocedural}, which present a
conservative abstraction to the polyhedral toolchain, 
and by (2) introducing more general predicates evaluated at runtime through 
fuzzy-array dataflow analysis~\cite{collard95fuzzy}.
In practice, conservative solutions mix well with the polyhedral
toolchain through a stubbing (a.k.a. blackboxing) mechanism and parallelism can
be expressed across irregular code regions. Unfortunately, this is not
sufficient as the decision to parallelize remains an all-or-nothing
compile-time decision performed at the granularity of the loop. In
contrast, \edt-based runtimes allow the expression of fine-grain parallelism down to the 
level of the individual instruction (overhead permitting).
Recently, new techniques, which allow for performing speculative and
runtime parallelization using the expressiveness of the polyhedral
model, have emerged~\cite{JimboreanCPML12}.

\cite{Oancea2012} make a very strong case that a dependence analysis
based on a \DAG\ of linear-memory array descriptors can generate lightweight
and sufficient runtime predicates to enable adaptive runtime
parallelism. This requires runtime evaluation of predicates and
results in significant speedups on benchmarks with difficult dependence structures. 
In their work, parallelism is still exploited in a fork-join model via the
generation of \omp\ annotations. 
One of the goals of task-based runtimes is to go beyond the
fork-join~\cite{openmp} and the bulk-synchronous~\cite{MPI-user-guide} models. 

\section{Approach}
This work is based on an automatic approach to program analysis and
transformation. From an analyzable sequential C specification, our algorithm performs
conversion to our intermediate representation and goes through the
following steps:
\begin{itemize}
\item Instance-wise dependence analysis with extensions to
  support encapsulated non-affine control-flow hidden within summary
  operations (a.k.a. blackboxes);
\item Scheduling to optimize a trade-off between parallelism, locality and
  other metrics in the program, using an
  algorithm~\cite{pat.locality}, which extends the work of~\cite{bondhugula08};
\item Non-orthogonal tiling of imperfectly nested loops with a
  heuristic, which 
  balances a model of data reuse, cache sizes and performance of streaming
  prefetches;
\item \edt\ formation from a tree representation of the tiled program;
\item Generation of dependences between \edts;
\item Code generation to a C++ runtime-agnostic layer (\ral).
\end{itemize}
The purpose of the \ral\ is to easily adapt to different runtimes.

\subsection{Notations}
%The polyhedral model is a mathematical abstraction to represent and reason
%about programs in a compact representation~\cite{Fea92,Pugh}. 
%We use the notations and conventions introduced by Girbal et al.~\cite{GV06},
%that we briefly recall here.
Our intermediate representation is based on a hierarchical dependence graph. 
The nodes of our graph are statements, which represent operations grouped
together in our internal representation. The unit of program analysis and
transformation is a statement; a statement $S$ can be simple or
arbitrarily complex (i.e., an external precompiled object), as long as it can be
approximated conservatively.
The edges of our graph are dependences as defined below.
An iteration domain for $S$, $D^S$, is an ordered multi-dimensional set of iterations. 
An instance of an iteration is written $i_S$.
The (lexicographic) order relation between iterations $i$ and $j$ is defined by $i\ll j$ iff $i$ occurs before $j$ in the program. 
By introducing $y$, the symbolic constant parameters of the program, an
iteration domain is the set $\{i_S \in D^S(y)\}$.
Operations to manipulate domains and their inverse include
projections to extract information along a subdomain; image by a function to
transform a domain into another domain; intersection to construct
the iterations that are common to a list of domains; and index-set splitting
to break a domain into disjoint pieces. 
Even at compile time, exact projection operations are often
prohibitively expensive, and we discuss the implications in section~\ref{subsec:dependence}.

A scheduling function $\Theta^S$ is a
linear affine function that partially reorders the iterations of $S$ in time. 
The order $\ll$ extends to time after scheduling is applied. 
In this context, a dependence $(T\rightarrow{S})$
is a relation\footnote{We use the notation $T\rightarrow{S}$ to express
  that $T$ depends on $S$} between the set of iterations of
$S$ and $T$. 
It conveys the information that some iteration $i^T\in D^T(y)$
depends on $i^S\in D^S(y)$
(i.e., they access the same memory location by
application of a memory reference) and that $i^S\ll i^T$ in the original
program. 
We write the set relation $\{(i^T,i^S)\in {\cal R}_{T\rightarrow S}(y) )\}$ or
${\cal R}_{T\rightarrow S}(y) $ to refer to the specific iterations of $T$ and $S$ that
take part in the dependence. 
The multigraph of statement nodes and dependence edges is referred to as the 
{\bf generalized dependence graph} (GDG). We write GDG=(V,E), the set of
vertices and edges in the graph respectively.

\subsection{Scheduling}
The R-Stream scheduler is a state-of-the-art parallelization tool,
which optimizes parallelism and locality in sequences of imperfectly nested
loops. An example of an earlier and more easily digestible algorithm
is presented in~\cite{bondhugula08}. Optimization is obtained by unifying the tilability conditions
expressed by~\cite{irigoin88} with scheduling techniques introduced by~\cite{feautrier92some}.
We briefly review the affine scheduling formulation. The input of the affine
scheduling problem is a GDG.
Following the standard conventions, $\phi_S$ is used to denote a 1-dimensional
affine schedule for statement $S$.
For each edge in the GDG, ~\cite{bondhugula08} writes:
\begin{equation}
   \delta(y) \ge \phi_{T}(i_T,y) - \phi_{S}(i_S,y) \ge 0,~~~~ 
        (i_T,i_S) \in {\cal R}_{T\rightarrow S}(y)
   \label{eq:delta}
\end{equation}
By combining all the dependences of the program, one forms a feasible linear
space that can be subject to various optimization problems.
The parametric affine form $\delta(y)$ can be interpreted
as the maximal dependence distance between any two schedules.
In particular, if $\delta(y)$ can be minimized to 0, then 
the solution $\phi$ is communication-free and is thus \emph{parallel}.
Similarly, if $\delta(y)$ can be minimized to a positive constant $c$, then
only local communication is needed and broadcast can be eliminated.
Bondhugula's iterative algorithm, allows finding independent solutions 
that are valid for the same set of dependence edges. This implies the induced
loops are permutable.

\begin{figure}[htbp]
\begin{algorithm}
1: \> \REPEAT \\
2: \>\>  Find as many independent solutions to (\ref{eq:delta}) as possible. \\
3: \>\>  \IF no solutions were found \THEN \\
4: \>\>\> Cut dependences between SCCs in the GDG. \\
5: \>\>  \ENDIF \\
6: \>\>  Remove from $E$ all edges satisfied in step (2). \\
7: \> \UNTIL $E = \O$;
\end{algorithm}
\caption{\label{fig:uday-algorithm} Bondhugula's algorithm.}
\end{figure}

Bondhugula's algorithm is shown in Figure~\ref{fig:uday-algorithm}.
A brief summary of the steps in the algorithm is in order:
\begin{itemize}
\item In step (2), we find as many solutions to (\ref{eq:delta}) as possible,
 while minimizing the coefficients to $\delta(y)$.  As the algorithm
 proceeds, edges from $E$ are removed.  In each iteration of step (2),
 we only consider the edges in $E$ that are remaining.
\item Steps (3)-(5) are triggered if step (2) fails to find a solution.
If so, dependences from different SCCs in the GDG are cut.  This has
the effect of fissioning the loops in different SCCs.  This edge cutting
can be done incrementally.
\item In step (6), all satisfied edges are removed from the graph.
An edge $e$ is {\em satisfied} by a solution $\phi$ iff
\[
  \phi_{\src(e)}(i,y) - \phi_{\dst(e)}(j,y) \ge 1, (i,j) \in 
      {\cal R}_{T\rightarrow S}(y) 
\]
\end{itemize}
Note that this edge removal step is not performed in step (2).  
This ensures that all solutions generated in step (2) are permutable, i.e.
valid for the same set of dependence edges.
The ability to extract general permutable loops is essential to scalable
generation of \edts.

\subsection{Scalability}
The expressiveness of the compositions of transformations allowed by the 
polyhedral model is also the source of scalability challenges. 
Traditionally, scalability challenges come in two flavors: 
scheduling computations and generating code for tiled loops.
In the context of \edts, a third challenge appears: 
the tractability of dependence computations between \edts.
The scalability of scheduling problem is out of the scope of this paper.
The scalability of code generation of tiled loops is of direct 
relevance to our work as we aim to generate \edts\ for multiple levels of 
hardware hierarchy. The example pseudocode sketched in 
section~\ref{sec:motivation} gives a first insight into the 
complexities involved when simultaneously optimizing parallelism and locality
in a polyhedral toolchain.
The constraints introduced for tiling can be expressed with simple loops 
that capture the constraints: 
$8t_1\le t_5\le 8t_1+7$, $16t_2\le t_6\le 16t_2+15$, $16t_3\le t_7\le 16t_3+15$
and $128t_4\le t_8\le 128t_4+127$.
However, the complete resulting code is significantly more complex and involves 
multiple min/max expressions as well as ceil and floor divisions.
The advantage is that optimal control-flow, in the number of loop iterations 
executed, is guaranteed~\cite{DBLP:conf/IEEEpact/Bastoul04}.
This is a common tradeoff between the complexity of the
representation (which is worst-case exponential in the number of induction
variables and symbolic parameters) and the quality of the generated
control-flow. 
In practice, tiling introduces additional induction variables at compile time
to specify different inter-tile and intra-tile iteration ordering and it is
rare to see more than two levels of tiling generated by pure polyhedral
solutions.
Parameterized tiling has been proposed as a solution to the 
scalability limitation in the polyhedral model for multiple levels of 
hierarchy~\cite{RKRSPLDI07,HBBCKNRS09}. Unfortunately, these techniques are 
non-linear and don't allow further optimizations to be carried within the
model. In particular, special considerations need to be taken into account 
to allow parallelism and even so, only a single level of imbalanced, 
wavefront-style parallelism is supported~\cite{DBLP:conf/cgo/BaskaranHTHRS10}.
A key innovation of this paper is based on the insight that \emph{the availability of \edt\ based runtimes changes this limitation in the
  context of parametric tiling.}

This work makes use of parameterized tiling. A tile in the polyhedral model 
 we propose to allow imperfect control-flow (which
may exhibit empty iterations) in order to achieve a more scalable
representation and the ability to generate multi-level code.
Techniques for symbolic Fourier-Motzkin elimination have been developed to 
reduce the potential overhead of empty iterations~\cite{DBLP:conf/cgo/BaskaranHTHRS10}
Symbolic runtime Fourier-Motzkin elimination is also used by~\cite{Oancea2012}.

%
%\nvnote{How to make the case for this point: This paper introduces a 
%compositional approach to ...}.
%
Lastly, an \edt-specific challenge lies in the tractable computation of 
dependences at compile-time and the overhead of their exploitation at runtime.
This is discussed in the next section.

\subsection{On Scalable Dependence Computation Between \edts}
\label{subsec:dependence}
The runtimes we target all require the programmer to specify dependences
between \edts\ to constrain the order of execution for correctness purposes. 
Dependence relations are exploited by the runtime to determine when a task is
ready and may be scheduled for execution.
An \edt-specific challenge lies in the tractable computation of 
dependences at compile-time and the overhead of their exploitation at runtime.

The requirements for dependence relations between \edts\ is significantly
different than for dependence analysis between statements.
Dependence analysis is only concerned with original program order
of statements and can be captured by the set.
$$
{\cal R}_{T\rightarrow S}(y) =\left\{(i^S, i^T)\in D^S\times D^T~|~i^S\ll
i^T,M_S[i^S]=M_T[i^T]\right\},
$$
where $M_S$ and $M_T$ are memory access functions (typically Read-After-Write
affine indices in the same array).
Array dataflow analysis~\cite{feautrier91dataflow} goes a step further and
takes into account all possible interleaved writes to only keep the true
producer-consumer dependences. A dataflow dependence is then written:
$$
{\cal F}_{T\rightarrow S}(y)={\cal R}_{T\rightarrow S}(y)-\left\{\Union_W\prod_{T\times S}\left({\cal R}_{T\rightarrow W}(y) \times{\cal R}_{W\rightarrow S}(y) \right)\right\},
$$
where $\prod$ is the projection operator from $T\times W\times S$ to 
$T\times S$.\footnote{In Feautrier, a parametric Integer Linear Programming
 (ILP) problem is solved and there is no need for a projection operator. The
formulation with set differences and the projector operator merely simplifies the
exposition of the problem.}
In the case of dependences between \edts\ these additional points must be
taken into account: 
\begin{itemize}
\item ordering must be computed on the transformed schedule which comprises tiling
  transformations, possibly at multiple levels,
\item multiple instances of a statement belong to the same tile. This is a
  projection operation that cannot be avoided when computing dependences
  exactly,
\item by virtue of exploiting parallelism, the ``last-write'' information
  becomes dynamic and introduces the need for sets of dependence relations.
\end{itemize}
A possible way to automatically specify these relations is to exactly
compute dependences between tasks at compile-time based on producer-consumer
relationships as has been proposed earlier~\cite{Baskaran2009}.
In this context, the following issues arise:
\par\noindent
\emph{1) Dependences may be redundant}. A straightforward dependence
algorithm considers producer-consumer relations on accesses to memory
locations. In the context of \edts\ without
special treatment to prune redundancies, all these dependences would be
generated translating into a high runtime overhead.
\par\noindent
\emph{2) Perfectly pruning dependences statically requires the static computation of
the ``last-write''} whose general solution is a Quasi-Affine Selection Tree 
(\quast)~\cite{feautrier88parametric}.
Computing this information exactly is often very expensive on original input
programs. After scheduling and tiling, the complexity of the ``last-write''
computation is further increased by the application of a projection operator
(multiple statement instances belong to the same tile instance).
An alternative is to compute ``covered dependences'' on the original
program~\cite{Pugh:1992} before scheduling and tiling and update them as
scheduling and tiling are applied. 
This approach still would require projection and redundancy
removal. Additionally, it is not expected to scale when multiple levels of
tiling are involved. 
 %\item  
\begin{figure}[htb] 
\centering
\includegraphics[width=.45\textwidth]{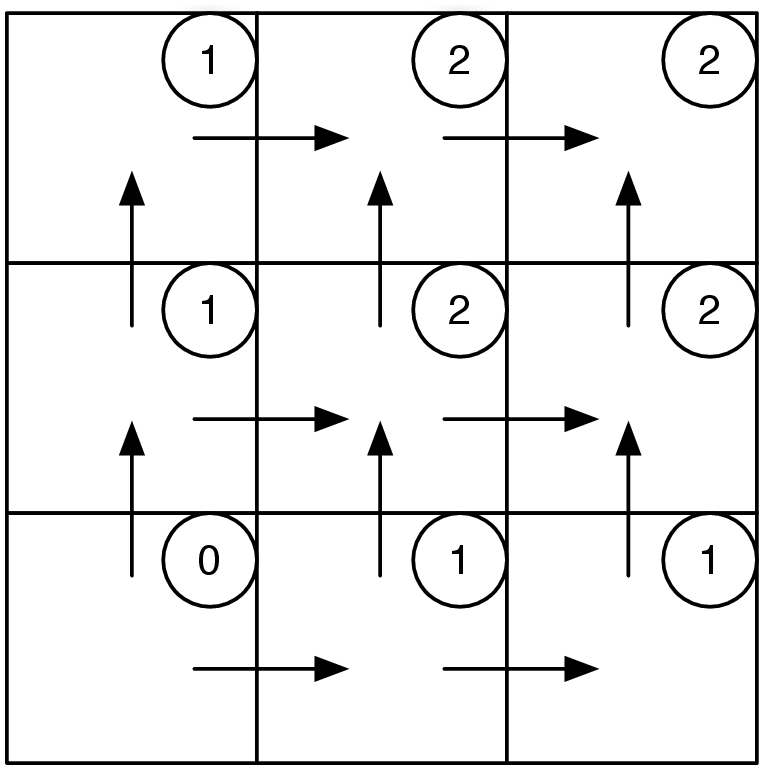}
\caption{\label{fig:non-convex-dependences}\edt\ antecedents graph}
\end{figure}

\par\noindent
\emph{3) Dependence relations between tasks are generally non-convex}, arising
from the projection of the dependence relation on a subspace. The projection
operator is non-convex.
In figure~\ref{fig:non-convex-dependences}, consider the possible dependence paths
between the origin $(i,j)=(0,0)$ and $(i,j)=(1,1)$. 
These correspond to the number of paths of Manhattan distance $2$ between
these points on a uniform 2-D grid. In particular, task
$(i,j)=(0,0)$ has no antecedents and can start immediately whereas
$(i,j)=(1,1)$ has $2$ antecedents. Task $(i,j)$ has $i\cdot j$
redundant dependences (i.e. the ``volume'' of the $i\times j$ region)
which reduces to $0$, $1$ or $2$ transitively unique dependences.  In
general, for one level of hierarchy, the number of these dependences
varies whether the task coordinates sit on the edge vertex, edge line
or interior of the 2D task space.  \cite{Baskaran2009}
handle this case by creating a centralized, sequential loop that scans
the Cartesian product of iteration spaces of source and destination
\edts\ for each single dependence. This mechanism incurs high
overhead.  In their experiments, tile sizes are kept larger to
amortize this overhead. However, if overprovisioning is to be
achieved, \edts\ have to be much smaller and the overhead of having
smaller \edts\ is very high in the work of Baskaran et al. We ran
experiments with the implementation of Baskaran et al. to confirm our
perceived limitations of their approach. Especially with benchmarks
where the loop nest is deeper, when the number of \edts\ is increased
with smaller tile sizes, the task graph with dependences incurs a huge
memory and performance overhead.

For all the reasons previously mentioned, we developed a new algorithm
based on loop properties rather than explicitly computing all possible
dependences between tasks.

\subsection{Tree Representation and \edt\ Formation}
After scheduling and tiling, the transformed program is represented as a tree
of imperfectly nested loops, similar to an abstract syntax tree (AST). 
Two main differences differences arise between a traditional AST and a tree of
loops in our model:
\begin{itemize}
\item our representation is oblivious to the effects of loop transformations
  (among which are peeling, loop shifting, parameter versioning and index-set
  splitting). 
  This is a particular strength of compositions of transformations in the
  model we adopted. Code generation deals with the intricacies of
  reforming the control-flow-efficient transformed loop nests~\cite{Bas04b}, 
\item subsequent loop transformations are further composable and preserve the
  independence of the representation with respect to complex control flow.
\end{itemize}
The tree structure is characterized by the integer ``beta vector'' that
specifies relative nesting of statement~\cite{GV06}:
\begin{itemize}
\item statements have identical first $d$ beta component if and only if they
  are nested under $d$ common loops. The bounds of the loops may be completely
  different for each statement,
\item as soon as the beta component differ, the loops are distributed,
  the order of the loops being consistent with the order of the beta component
\end{itemize}
Each node in the tree therefore corresponds to a loop and has a loop type
associated with it. To uniformize the \edt\ extraction algorithm, we also
introduce a root node in the tree that does not 
correspond to any loop but is the antecedent of all nodes.  
The algorithm in Figure~\ref{fig:edt-formation} performs a 
breadth-first traversal on the tree structure induced by the beta vectors and
marks nodes.
The algorithm forms sequences of perfectly nested consecutive loops with
compatible types for dependence inference purposes (see
section~\ref{sec:deptypes}). In particular, permutable loops of the same band
can be mixed with parallel loops. However, permutable loops belonging to
different bands cannot be mixed in the current
implementation.

\begin{figure}[htbp]
\begin{algorithm}
1. \> Mark the root node of the tree\\
2: \> \REPEAT \\
3: \>\>  Let N be the next node in the BFS traversal \\
4: \>\>  \IF N is at tile granularity \OR \\
5: \>\>\>    N is user-provided \THEN mark N  \\
6: \>\>  \ELSE \IF N is sequential \THEN mark N \\
7: \>\>  \ELSE \IF N is has siblings \THEN mark N \\
8: \>\>  \ELSE \IF N is permutable \AND \\
9: \>\>\>  N is in a different band than parent(N) \AND\\
10: \>\>\>  parent(N) is not marked \\
11: \>\> \THEN \\
12: \>\>\>  mark N \\
13: \>\>  \ENDIF \\
14:\> \UNTIL tree traversed;\end{algorithm}
\caption{\label{fig:edt-formation} \edt\ formation algorithm}
\end{figure}
At the moment, we support two strategies. The default approach consists of
stopping traversal when the granularity of a tile is reached. This creates
\edts\ at the granularity of tiles. The second strategy lets the user decide
which nodes should be marked and ignores tile granularities. The algorithm
introduces the remaining nodes necessary to accommodate changes in permutable
bands as well as sequential loops and imperfectly nested loops that would
require too many dependences and that we handle in a special way (see 
section~\ref{sec:deptypes}). 

Once the tree is marked, one compile-time \edt\ is formed for each marked
\emph{non-root} node. This proceeds as follows:
\begin{enumerate}
\item pick a new unique id for the current \edt, 
\item determine the start and stop level for this \edt. The start level is the 
  level of the first marked ancestor, the stop level is the level of the node,
\item filter the statements nested below the node and attach them to the
  current \edt,
\item form the iteration domain of the \edt\ as the union of the domains of
  all its statements.
\end{enumerate}
One ends up with a tree of \edts, {\bf whose coordinates in a multi-dimensional tag
space are uniquely determined by loops $[0,stop]$}. Coordinates $[0,start)$
are received from the parent \edt, coordinates $[start, stop]$ are determined
locally from loop expressions. The combination of an \edt\ id and its
coordinates uniquely identify each \edt\ instance.

\begin{figure*}[htb]
\centering
\includegraphics[width=0.90\textwidth]{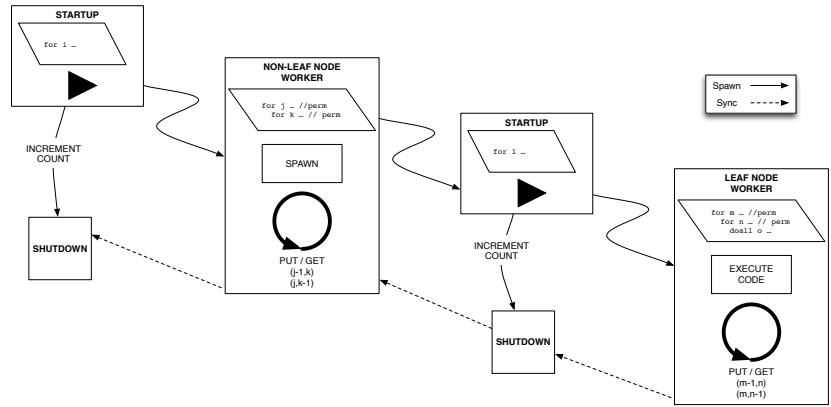}
\caption{\label{fig:hierarchy-big-picture} Global organization of
  \edts\ spawning and synchronization.}
\end{figure*}

The code generation process creates three different runtime \edts\ for each
compile-time \edt:
\begin{enumerate}
\item a \Startup\ \edt\ spawns \Worker\  \edts\ asynchronously and sets up a counter to
determine when a subgroup of \Worker\  \edts\ is complete, 
\item a \Worker\  \edt\ either further decomposes into additional levels of \edts\ or
performs work if it is a leaf in the tree, and 
\item a \Shutdown\ \edt\ acts as a synchronization point for tasks generated
  in step 1. 
\end{enumerate}
Vertically, across steps 1 - 3, counting dependences capture hierarchical
async-finish relations and are used to implement hierarchical synchronization points. 
Horizontally, within step 2, point-to-point dependences provide load-balancing
support.
Figure~\ref{fig:hierarchy-big-picture} illustrates the organization of
spawning and synchronizations across the three types of \edts.
This overall structure is further described in section~\ref{sec:asynfinish}. 

\subsection{Dependence Specification With Loop Types Information}
\label{sec:deptypes}
When performing dependence analysis, one obtains a direct relation between
iterations in the source and target domains. 
These relations should be cheap to determine at compile time (i.e. not require
projections), should evaluate at runtime in a distributed, asynchronous mode
and should not require iterating over a high-dimensional loop nest at runtime.
To avoid iterations over a high-dimensional loop nest, we adopt a
decentralized view of the problem using a \emph{get-centric} approach in which
an \edt\ queries its predecessors  whether they have finished executing.
This minimizes the number of \emph{put}s in a concurrent hash table, which are
notoriously more expensive than gets.  

Individual dependence relations between statements are generally
non-invertible. Consider the relation $[i, i]\rightarrow[i,
  j]$~\footnote{With our convention this reads: iteration $[i,i]$ depends
  on iteration $[i,j]$}. Forming
the inverse relation requires a projection and gives rise to (many)
dependences at runtime. 
A key insight of this work is that although individual dependence relations
between statements may require projections, \emph{aggregate dependence
  relations} between \edts\ may not.
Loop types exhibited by the scheduler and described in this section allow
the scalable computation of these aggregate dependences.

Parallel loops are the simplest type of dependence relation in the program:
they carry no dependence. As a consequence, no special conditional needs to be
evaluated at runtime.

A permutable band of loops over induction variables $(i_1,\dots,i_n)$
has only forward dependences. These can always be expressed conservatively by
the set of $n$ invertible relations: 
$$\{[i_1, i_2, ..., i_n]+e_k\rightarrow[i_1, i_2, ..., i_n], k\in[1,n]\},$$ 
where $e_k$ is the canonical unit vector in dimension $k$.
In order to infer dependences for a nest of permutable loops, each task
determines from its coordinates in the task space whether it is a boundary
task or an interior task and which other tasks it depends on.
For the purpose of illustration, we consider 3D tasks (i.e., loops
\verb|(i, j, k)| are loops across tasks or inter-task loops). 
The code shown in Figure~\ref{fig:permutable} exploits dependences of distance $1$ along each dimension
(i.e., Boolean expressions are formed plugging \verb|i-1|, \verb|j-1|, and
\verb|k-1| into the expression of the loop bounds). 
For each permutable loop dimension, we introduce a condition to determine
whether the antecedent of the task along that dimension is part of the interior
of the inter-task iteration space. When the condition evaluates to \verb|true|, the
task must wait (i.e. get) for its antecedent to complete.
Due to space considerations we only show the interior\_1 computation;
interior\_2 and interior\_3 are omitted.

Sequential is the most restrictive type for loops. It imposes a fully
specified execution order on the current loop with respect to \emph{any loop nested
  below it}. To visualize these effects, consider the code in
figure~\ref{fig:sequential} where the function f reads a portion of array A
such that the dependence structure is \verb|(seq,doall,seq,doall)|.
Suppose that a task has the granularity of a single \verb|(t,i,j,k)| iteration
of the innermost statement. 
\begin{figure}[htb] 
\begin{rulett}
\scriptsize{
for (t=1; t<T; t++) \{
  doall (i=1; i<N-1; i++) \{
    for (j=1; j<N-1; j++) \{
      doall (k=1; k<N-1; k++) \{
        A[t][i][j][k] = f(A);
\}\}\}\}
}
\end{rulett}
\caption{\label{fig:sequential}Effect of sequential loops}
\end{figure}
The dependence semantic from loop \verb|t| is that \emph{any\/} task
\verb|(t,i,j,k)| depends on \emph{all\/} of its antecedents
\verb|(t-1,*,*,*)|.
Similarly from loop \verb|j|, \emph{any\/} task \verb|(t,i,j,k)| depends on
\emph{all\/} of its antecedents \verb|(t,i,j-1,*)|.
If all dependence relations were to be exposed to the runtime, the \verb|t|
loop would require $N^6$ dependences which is prohibitive. 
The naive solution gets even worse when multiple
levels of hierarchy are introduced: if each loop \verb|i|, \verb|j|, \verb|k| were split in $2$ equal
parts, each task's granularity would be reduced by a factor of $2^3$. The naive
approach would yield an increase in number of dependences by a factor of $2^6$
and is clearly not a viable alternative.
\begin{figure}[htb] 
\begin{rulett}
\scriptsize{
// Pseudo-code to spawn tasks asynchronously
for (i=\(\lceil\frac{-N-15}{16}\rceil\);i<=\(\lfloor\frac{T-3}{16}\rfloor\);i++) \{
 for (j=max(i,-i-1);
      j<=min(\(\lfloor\frac{-8*i+T-2}{8}\rfloor\),\(\lfloor\frac{8*i+N+7}{8}\rfloor\),\(\lfloor\frac{T+N-2}{16}\rfloor\));j++) \{
  for (k=max\((0,\lceil\frac{i+j-1}{2}\rceil,\lceil\frac{16*j-N-14}{16}\rceil)\);
    k<=min\((\lfloor\frac{T+N-2}{16}\rfloor,\lfloor\frac{16*j+N+14}{16}\rfloor,\lfloor\frac{8*i+8*j+N+15}{16}\rfloor)\);k++) \{
     spawn\_task\_async(tag(i, j, k));

// Input: tag t
int i=t.i, j=t.j, k=t.k;
// Use enclosing loops' upper and lower bounds to 
// determine whether the task sits on a boundary
bool interior\_1 = 
  (({\bf i-1} >= \(\lceil\frac{-N-15}{16}\rceil\)) && ({\bf i-1} <=  \(\lfloor\frac{T-3}{16}\rfloor\)))                   &&
  ((j >= ({\bf i-1},-i))                                           && 
   (j <= min(\(\lfloor\frac{-8*({\bf i-1})+T-2}{8}\rfloor\),\(\lfloor\frac{8*({\bf i-1})+N+7}{8}\rfloor\)),\(\lfloor\frac{T+N-2}{16}\rfloor\))) &&
  ((k >= max\((0,\lceil\frac{({\bf i-1})+j-1}{2}\rceil,\lceil\frac{16*j-N-14}{16}\rceil)\))                 && 
   (k <= min\((\lfloor\frac{T+N-2}{16}\rfloor,\lfloor\frac{16*j+N+14}{16}\rfloor,\lfloor\frac{8*({\bf i-1})+8*j+N+15}{16}\rfloor)\)));
if (interior\_1) wait\_task(tag({\bf i-1}, j, k));  // a.k.a get
//   \edt\ body
signal\_task\_finished(tag(i, j, k));         // a.k.a put
}\end{rulett}
\caption{Antecedent determination in permutable
  loop nest}
\label{fig:permutable}
\end{figure}
A simple idea consists in generating a 1D fan-in/fan-out task similar to a
\verb|tbb::empty_task|. This has the effect of reducing the ``Cartesian
product'' effect of dependence relations. 
The dependence semantic becomes: \emph{any\/} task \verb|(t,i,j,k)| depends on \verb|sync(t)|
and \verb|sync(t)| depends on \emph{all\/} tasks \verb|(t-1,i,j,k)|.
The number of dependences reduces to $2N^3$ which can still be
too high. 
% but still increases by
%$O(n^3)$ as multiple levels of granularity are introduced.
The approach we exploit for sequential loops is based on \emph{hierarchical separation
of concerns}: a sequential loop generates an additional level of
hierarchy in the task graph, effectively acting as a tbb::spawn\_root\_and\_wait. 
To accommodate this separation of concerns, we generate three different runtime
\edts\ for each compile time \edt.

\paragraph{Towards More Flexible Semantics}
\label{sec:flexible}
Without deeper compiler introspection, the loop semantics we described may
sometimes be too conservative. 
In figure~\ref{fig:conservative}, we provide two toy examples in which
conservativeness reduces parallelism.
\begin{figure}[htb] 
\begin{rulett}
\scriptsize{
for (t=1; t<T; t++) \{                                        for (t=1; t<T; t++) \{       
  doall (i=1; i<N-1; i++) \{                                    doall (i=1; i<N-1; i++) \{ 
    A[t+1][i] = C0*A[t-1][i];                                     A[t][i] = C0*A[T-t][i]; 
\}\}\}\}                                                         \}\}\}\}                      
}
\end{rulett}
\caption{\label{fig:conservative} Conservative dependence distance (left) and
  index-set (right)} 
\end{figure}
In the first case, dependence distances of length $1$ are too conservative.
Dependence distances of length $2$ enable twice as many tasks to be executed
concurrently. More generally, for a set of dependences with constant
distances, the greatest common divisor of the distances gives the maximal
amount of parallelism.
In the second case, the \verb|t|-loop could be cut (by index-set-splitting) in $2$
portions that do not contain any self-dependence. More generally, for a set of
dependences, the transitive closure of dependence relations is needed to fully
extract the maximum available parallelism~\cite{Griebl99indexset}.

The two simple cases presented can be captured in a simple framework
implementing the technique illustrated in this section.  The first case can be dealt with by
specifying a \emph{dependence relation between inter-task indices\/}
that generalizes the relation illustrated in
figure~\ref{fig:permutable} (which only considers $\left\{\mbox{\tt
  t-1}\rightarrow \mbox{\tt t}\right\}$, $\left\{\mbox{\tt
  i-1}\rightarrow \mbox{\tt i}\right\}$ and $\left\{\mbox{\tt
  j-1}\rightarrow \mbox{\tt j}\right\}$).  The second case can be
solved by augmenting the computations of the Boolean expressions
(\verb|interior_1|, \verb|interior_2|, and \verb|interior_3|) to include additional
filtering conditions.  In other words, \emph{the effect of
  index-set-splitting\/} is applied on the Boolean computation only and
\textbf{not on the iteration domain of the statements, as is
  customary}. The consequence is that the complexity of control-flow
resulting from index-set splitting will not contaminate the iteration
domains of the statements as well
as potentially non-convex dependences (replaced by the Boolean
evaluations in figure~\ref{fig:permutable}).

The cases discussed above are simple illustrations and multiple difficulties
arise when both cases come together to provide additional parallelism. 
Tradeoffs will exist between the tractability of the static
analysis involving the computation of transitive closures of the different
sets of dependences~\cite{Griebl99indexset}, scalability to deep levels of hierarchy,
and impact of the information derived on the amount of parallelism in the
program. It will be of paramount importance to focus efforts on cases that
occur in practice in applications of relevance to Exascale.

\subsection{Runtime Agnostic Layer}
\label{sec:ral}
The \ral\ comprises of a set of C++ templated classes to build expressions
evaluated at runtime, along with an API that our compiler targets. 
The API aims at being a greatest common denominator for features across
runtimes. 
Since our compiler does not support C++ natively, we hide some of the
implementation behind C macros. Note that in order to support \swarm, C macros
are extensively needed.
The central element is the templated TaskTag which encapsulates the tuple
holding the coordinates of the \edt\ in the tag space.
Specialized tuples for each runtime derive from this TaskTag and optionally
extend it with synchronization constructs to implement async-finish.
TaskTags are passed along with \edts\ as function parameters in \swarm,
\ocr\ and \cnc.

\subsubsection{Templated Expressions}
We implemented an approach based on C++ template expressions to capture complex
loop expressions and dynamically evaluate inverse dependence relations.

\begin{figure}
\begin{grammar}
\small{
<linear-expr>      ::= <number>
\alt <induction-term>
\alt <parameter>
\alt <number> `* ' <linear-expr>
\alt <linear-expr> `+ ' <linear-expr>
\alt <linear-expr> `- ' <linear-expr>

<expr>             ::= <linear-expr>
\alt `MIN(' <expr> `, ' <expr> `)'
\alt `MAX(' <expr> `, ' <expr> `)'
\alt `CEIL(' <linear-expr> `, ' <number> `)'
\alt `FLOOR(' <linear-expr> `, ' <number> `)'
\alt `SHIFTL(' <linear-expr> `, ' <number> `)'
\alt `SHIFTR(' <linear-expr> `, ' <number> `)'

<comp-expr-list>   ::= <comp-expr> `, ' <comp-expr-list> | <comp-expr> 

<multi-range>      ::= <range> `;' <multi-range> | <range> 
}
\end{grammar}
\caption{\label{grammar} Range grammar}
\end{figure}

Figure~\ref{grammar} encodes the multi-dimensional ranges that can be
generated by our method. Operations on these ranges take tuples as
input and return Booleans or other tuples. Supported operations include
evaluation of the expression at a tuple, comparisons at a tuple and computations
of the minimum and maximum given a tuple range (bounding box computation).
These expressions are used as described below, following
Figure~\ref{fig:permutable}. 
First a tuple of ``terms'' is created that encapsulates the induction
variables (\verb|t1|, \verb|t2|, \verb|t3|) and parameters (\verb|T|, \verb|N|). 
Then, templated expressions \verb|p1|, \verb|p2|, and \verb|p3| are
declared that capture the lower and upper bound expressions governing the
iterations of the terms.
Lastly, based on the lower and upper bound expression, runtime dependences are
generated using a templated construct that dynamically captures the
non-convex Boolean evaluations from figure~\ref{fig:permutable}.
These expressions are oblivious to the complexity of the loop expressions,
which can become a severe bottleneck in a polyhedral \ir\ based on dual
representation of constraints and vertices.
The tradeoff is the runtime overhead for constructing and evaluating the
expression templates. 
Experiments with vtune and particular efforts in keeping this overhead
low by using C++11's \verb|constexpr| types and declaring the expressions
static, show an overhead below 3\% in the worst cases we encountered.

\subsubsection{\edt\ Code Generation}
We extended the \cloog\ code 
generator~\cite{Bas04b} to walk a tree of \edts\ in a recursive, top-down
traversal. Each \edt\ is generated in its own separate file. These are later
compiled independently with gcc and linked with the runtime library to produce
the final executable.
\Shutdown\ \edts\ do not require special treatment, they always consist in
the same code, only parametrized by the \TaskTag.
Each \Startup\ and \Worker\ \edt\ is parametrized by a start and a stop level.
Until the start level, three behaviors can be observed:
 (1) induction variables and parameters are retrieved directly from the
\edt\ tag by using the overloading of TaskTag::operator=, (2) loops are 
forcibly joined by forming the union of their domains and (3) loops are
forcibly emitted as conditionals.
Between the start and stop levels, the code generation process follows the
normal behavior of \cloog, separating statements and generating loops and
conditionals in each subbranch of the code tree.
After the stop level, behavior depends on the type of \edt:
\begin{itemize}
\item for a \Startup, we generate the counting variable increment in the first loop
  and the spawning of \Worker\ \edts\ in the second loop,
\item for a non-leaf \Worker, we generate code to recursively spawn a
  \Startup,
\item for a leaf \Worker, the computations and communications corresponding to
  the actual work are generated.
\end{itemize}

\subsubsection{Runtimes Discussion}
Concurrent Collections (\cnc) is a high-level coordination language that lets 
a domain expert programmer specify semantic dependences in her program without worrying
about what runs where and when. %Later, an optimization expert can optimize the
%program thus achieving separation of concerns. 
\cnc\ has a task-graph model that is \emph{implicit} in the program
representation by using hash tables.  
Intel-\cnc\ ships with a work-stealing runtime whose default scheduler is
built on top of the scheduler provided by Intel's Threading Building
Blocks(\tbb)~\cite{TBB}. \tbb\ was strongly inspired by
\stapl~\cite{RauchwergerAO98}.
%\cnc\ also allows a separate tuning layer to influence the scheduler
%to specify where a step should run or which steps produces / consumes
%which item. We do not take advantage of these features in this work.
%Intel-\cnc\ is written in C++ on top of Intel's TBB API. 
\cnc\ uses 
tbb::concurrent\_hashmap to implement step and item collections.
A step is a C++ object that implements an execute method; it represents a
scheduled unit of execution. The \cnc\ scheduler decides at runtime which
step::execute methods are called on which hardware thread and on which
processor. 
%In a possible mode of execution, \cnc\  has a step queue per hardware
%thread; when a thread is done with a step it takes the next ready step from
%its queue. If the queue is empty, it randomly steals~\cite{Dinan:2009} half of
%the steps from the queue of another thread. 
This step::execute method takes a step tag reference and a context reference.
A step becomes available when an associated step tag is put in the proper step
collection. A step may perform multiple gets and puts from/to item collections.
Item collections act as dataflow dependence placeholders. 
By default, a \cnc\ get is blocking. If it fails, control is given back to
the scheduler which re-enqueues the step to await the corresponding tag put.
Once that put occurs, the step restarts. 
In the worst-case scenario, each step with N dependences could do N-1 failing
gets and be requeued as many times. Additionally, on a step suspension, the
gets are rolled back.
Performing all gets of a step before any put offers determinism
guarantees~\cite{CnCZoran}.
%To avoid prohibitive overheads from multiple dependences, \cnc\ supports 2
%mechanisms for pre-specifying dependences. The first mechanism consists in
%implementing a tuner class with a  ``depends'' function. The depends function
%is called at the time a tag is inserted into a tag collection (i.e. the first
%time the runtime ``sees'' a particular step instance). 
%The second mechanism consists in performing calls to non-blocking
%``unsafe-get'' methods. Unsafe gets may return false if the object of the get
%is not yet available. A subsequent call to flush\_gets triggers an exception
%which undoes the gets and reschedules the step to wait for all 

ETI's SWift Adaptive Runtime Machine (\swarm) \cite{swarm} is a low-level parallel computing
framework that shares similarities with \cnc. Additionally, \swarm\ handles
resource objects and allows active messages and continuation passing style.
In this work we only exploit the features that are common to \cnc. 
\swarm\ is a C API that makes extensive use of pre-processor macros.
In \swarm, an \edt\ is declared as a C macro and scheduled
into the runtime by calling the swarm\_schedule function. An \edt\ accepts a
context parameter THIS and an optional parameter INPUT that come in the form
of pointers. \swarm\ allows more complex behaviors where a parent \edt\
specifies a NEXT and NEXT\_THIS parameter to allow chaining of multiple
\edts. \swarm\ also allows an \edt\ to bypass the scheduler and dispatch
another \edt\ immediately using swarm\_dispatch.
The tagTable put and get mechanisms in \swarm\ are fully non-blocking.
It is the responsibility of the user to handle the synchronization properly,
re-queue \edts\ for which all gets did not see matching puts and terminate the
flow of execution for such \edts.
\swarm\ presents a lower-level runtime and API and allows many low level
optimizations. 

The Open Community Runtime (\ocr~\cite{OCR}) is a third runtime system based
on \edts\ and work-stealing principles. \ocr\ represents the task graph
explicitly and does not rely on tag hash tables. In \ocr\ different objects can
be specified as ``events'', whether they represent \edts, blocks of data
(``datablocks'') or synchronization objects. 
\ocr\ does not natively rely on a tag space. Instead, when an \edt\ is
spawned, all the events it depends on must have already been created by the
runtime and must be passed as dependence parameters to the \edt. 
By contrast, in \cnc\ and \swarm, when a get is performed, the corresponding
hash table entry can be viewed as a type of ``synchronization future.''
There is effectively a race condition between the first get, the subsequent
gets and the first put with a given tag. Additionally, mapping to a tag tuple
to an event is necessary to create the synchronizations.
Without a hash table, \ocr\ requires the pre-allocation of a large
number of synchronization events (as is demonstrated in the Cholesky example
that ships with \ocr).
We chose to implement a \emph{prescriber} in the \ocr\ model to solve this
race condition. Puts and gets are performed in a
\emph{tbb::concurrent\_hash\_map} following the
\cnc\ philosophy.~\footnote{Incidentally this is the same approach as chosen
  by the \ocr\ team when implementing \cnc\ running on top of \ocr.}
This \prescriber step is completely oblivious to the compiler and is fully
handled by the \ral. In our targeting \ocr, each \worker\ \edt ~ is dependent on
a \prescriber\ \edt ~which increases the total number of \edts.
Lastly, \ocr\ is the only of the three systems to support hierarchical
async-finish natively via the use of a special ``finish-\edt''.

\cnc, \swarm\ and \ocr\ run on both shared and distributed memory systems
(\ocr\ extensions in progress for distributed systems). In this work, we only
target the shared memory implementations.

\subsection{Runtime Support for Hierarchical Async-Finish}
\label{sec:asynfinish}
In this section we discuss the support for hierarchical async-finish tasks in
\ocr, \swarm\  and \cnc. Our tool automatically generates \edts\ that
conform to such a hierarchical execution model from sequential input code.

\begin{figure}[htb] 
\centering
\includegraphics[width=0.90\textwidth]{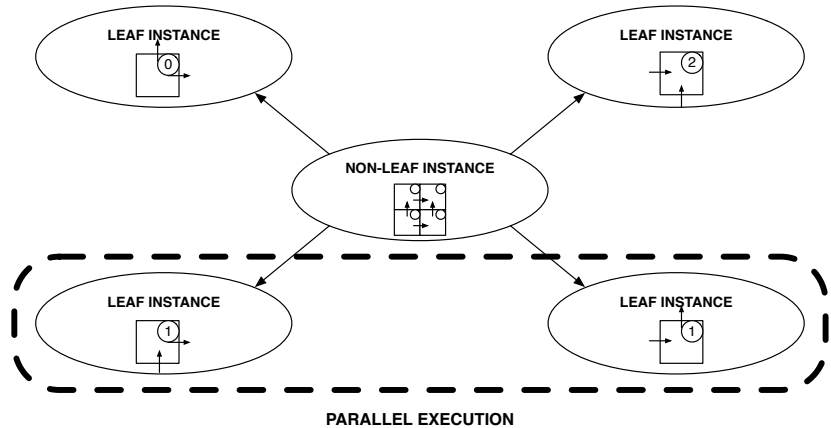}
\caption{\label{fig:hierarchy}Parallelism among hierarchical \worker \edts.}
\end{figure}
Figure~\ref{fig:hierarchy} illustrates parallelism across
hierarchical \worker\ \edts. \\
\worker\ instances in the non-leaf worker (center
circle) are connected by point-to-point dependences. Within each top-level
\worker, bottom-level \worker\ are spawned, themselves connected by
point-to-point dependences. Instances not connected by dependences
(i.e. the unordered bottom-left and bottom-right instances in this small example)
can be executed in parallel by the runtime. This is a coarse level of
parallelism. Additionally, within each leaf worker, finer grained parallelism
is also exploited by the runtime.  

\ocr\ natively supports hierarchical async-finish by virtue of the 
``finish \edt.'' \ocr\ also provides ``latch'' objects that can be used to
emulate this feature like in \swarm\ (see below).
The other two runtimes do not currently provide native support
and we construct a layer of emulation that our source-to-API compiler targets
automatically.

\swarm\ natively supports ``counting dependence'' objects which are similar to
\ocr\ latches . 
%The underlying \swarm\ lock-free, wait-free implementation is optimized to minimize
%hotspots and uses a reduction tree to efficiently perform parallel increments.
We use this feature as follows: 
\begin{itemize}
\item within each \Startup\ we generate code which determines how many
  \Worker\  are spawned. A swarm\_Dep\_t object is allocated and default
  initialized to this value,
\item when both the counter and the counting dependence are ready, 
  a \Shutdown\  is chained to await on the dependence object with the
  associated count value. When the dependence count reaches zero, the
  \Shutdown\  is awoken,
\item a pointer to the swarm\_Dep\_t object is passed as a parameter into the
  tag of each \Worker\  instance. At this point, the current instance of
  \Startup\  can spawn all its \Worker s,
\item when multiple levels of hierarchy are involved, each instance of a leaf
  \Worker\  satisfies the dependence to the \Shutdown\  spawned by
  their common enclosing \Startup,
\item non-leaf \Worker\  relegate the dependence satisfaction to the
  \Shutdown\  spawned by the same \Startup\  instance. 
\end{itemize}
\Shutdown\  satisfies the counting dependence of their caller, up until
the main \Shutdown\  which stops the runtime. 
%This somewhat complex
%behavior is induced by the ability of \swarm\ to notify an \edt\ when the
%dependence count reaches zero. 
%More specifically, the intricacy comes from the declaration and chaining being
%performed by the caller, whereas the satisfaction is performed by the callee.
%One difference in the current \ocr\ implementation is that \swarm\  dependences
%can be incremented and decremented by arbitrary values while \ocr\ dependences
%only can only be incremented and decremented by one \footnote{This could be a
% synchronization bottleneck for \ocr\ if it did not support async-finish natively.}.

\cnc\ does not natively support async-finish or even counting dependences. A
reduction operator is under development but yet unavailable to us at the time of
this writing. We settle for a simple solution using a C++11 \verb|atomic<int>|. 
Each \Worker, upon completion performs an atomic decrement of the shared
counter. Like for \swarm, the counter is constructed and passed by calling
\Startup. 
Unlike \swarm, the ability to notify the \Shutdown\ on the event that the
counter reaches zero is missing.  
To perform this synchronization in \cnc, a \Shutdown\ performs a get
of an item that is only put in the corresponding item collection by the unique
\Worker\  \edt\ that decrements the counter to zero (i.e. the dynamically
``last'' one).
Unlike \swarm\ and \ocr, which provide their own mechanisms, this emulation
relies on the item collection (a hashtable) to perform the signaling. 
However, accesses this hashtable are very rare: only the last \Worker\ and the
associated \Shutdown\ write and read it respectively.
 
%The lack of a support for counting dependences in \cnc\  makes the programming
%model somewhat easier than in \swarm.

%Key difference between \cnc\ and \swarm:
%In \cnc, satisfyOnce does an atomic decrement and the last to decrement also
%does a tag put to the enclosing \Shutdown.

%\begin{table*}
%\begin{tabular}{cccc}
%Feature / Runtime & \cnc\ & \ocr\ & \swarm\ \\ 
%\hline \\
%Dependence specification & 1-sided put/get with tags & 2-sided dependences & 1-sided put/get with tags\\
%\hline \\
%Dependence data structure & TBB concurrent hashmap & ? & custom hashmap\\
%\hline \\
%\hline \\
%Synchronization & atomic counter + tags & ? & \\
%\hline \\
%Programming model & C++ API & low-level API & low-level API + object model\\
%\hline \\
%Ease of Programming & +++ & + & x\\
%\hline
%\end{tabular}
%\end{table*}
%\caption{\label{fig:}Runtime features discussed in this paper}

%
%\subsubsection{Recursive Hierarchical Decomposition}
%Need for adaptivity
%
%To each worker \edt we associate 

\begin{table*}
%\hspace{-10mm}
  {\scriptsize
    \begin{center}
%    \begin{tabular}{rl}
      \begin{tabular}{|l||l|llllll|}
  \hline
  Benchmark & \edt\ version & 1 th. & 2 th. & 4 th. & 8 th. & 16 th. & 32 th.\\
  \hline
  \hline
 & \cncdepend & 2.53 & 4.51 & 6.52 & 6.65 & {\bf 7.24} & 5.27\\
\divergencethreedoneiteration & \cncsyncget & 2.54 & 4.41 & 5.87 & {\bf 6.91} & 6.72 & 5.09\\
 & \cncunsafeget & 2.82 & 2.82 & 2.62 & 2.60 & 2.36 & 2.18\\
\hline
 & \cncdepend & 1.19 & 2.21 & 4.06 & 7.19 & {\bf 10.85} & 10.03\\
\fdtdtwod & \cncsyncget & 1.12 & 2.16 & 3.98 & 6.88 & 8.44 & 7.94\\
 & \cncunsafeget & 1.17 & 2.23 & 4.31 & 7.94 & {\bf 12.20} & 10.73\\
\hline
 & \cncdepend & 0.98 & 1.90 & 3.04 & 5.27 & {\bf 8.45} & {\bf 10.85}\\
\gaussseideltwodfivepointsenlarged & \cncsyncget & 0.81 & 1.56 & 2.90 & 4.58 & 4.86 & 4.45\\
 & \cncunsafeget & 0.84 & 1.65 & 3.13 & 5.75 & 6.26 & 5.25\\
\hline
\ & \cncdepend & 0.98 & 1.96 & 3.30 & 5.67 & 9.42 & {\bf 13.72}\\
\gaussseideltwodninepointsenlarged & \cncsyncget & 0.91 & 1.76 & 3.34 & 5.94 & 8.30 & 7.23\\
 & \cncunsafeget & 0.95 & 1.85 & 3.61 & 6.56 & {\bf 11.29} & 8.96\\
\hline
 & \cncdepend & 1.82 & 3.63 & 6.54 & 11.74 & 21.89 & 32.14\\
\gaussseidelthreedtwosevenpointsenlarged & \cncsyncget & 1.78 & 3.61 & 6.93 & 12.87 & 24.78 & {\bf 37.41}\\
 & \cncunsafeget & 1.81 & 3.56 & 6.88 & 12.89 & 24.64 & {\bf 37.77}\\
\hline
& \cncdepend & 1.60 & 3.19 & 5.02 & 8.46 & 14.80 & 21.56\\
\gaussseidelthreedsevenpointsenlarged  & \cncsyncget & 1.52 & 3.05 & 5.84 & 10.81 & 20.58 & {\bf 33.06}\\
 & \cncunsafeget & 1.54 & 3.06 & 5.87 & 10.93 & 20.40 & {\bf 33.89}\\
%\hline
%\gradientthreedoneiteration & \cncdepend & 1.17 & 2.10 & 2.90 & 3.84 & {\bf 3.98} & 3.30\\
% & \cncsyncget & 1.17 & 2.41 & 3.04 & 3.85 & {\bf 4.13} & 3.69\\
% & \cncunsafeget & 1.36 & 1.28 & 1.27 & 1.33 & 1.20 & 1.11\\
\hline
 & \cncdepend & 4.28 & 7.57 & 12.79 & 21.43 & {\bf 28.73} & 23.39\\
\jactwod & \cncsyncget & 3.53 & 6.71 & 11.39 & 14.17 & 13.34 & 13.22\\
 & \cncunsafeget & 3.81 & 7.28 & 13.77 & 23.93 & {\bf 26.21} & 23.92\\
\hline
 & \cncdepend & 2.05 & 4.01 & 5.96 & 8.92 & 13.51 & {\bf 16.88}\\
\jacobidoubletwodfivepointsenlarged & \cncsyncget & 1.57 & 2.96 & 5.49 & 8.48 & 8.80 & 8.16\\
 & \cncunsafeget & 1.58 & 3.15 & 5.98 & 10.76 & {\bf 16.35} & 10.17\\
\hline
 & \cncdepend & 1.46 & 3.06 & 5.56 & 10.24 & 17.54 & 17.78\\
\jacobidoubletwodninepointsenlarged & \cncsyncget & 1.51 & 2.86 & 5.47 & 9.73 & 15.09 & 15.76\\
 & \cncunsafeget & 1.55 & 2.96 & 5.58 & 10.72 & {\bf 19.13} & {\bf 18.80}\\
\hline
 & \cncdepend & 2.42 & 4.75 & 8.72 & 15.66 & 25.69 & 28.10\\
\jacobidoublethreedtwosevenpointsenlarged & \cncsyncget & 2.41 & 4.90 & 9.20 & 16.77 & 31.64 & {\bf 34.96}\\
 & \cncunsafeget & 2.39 & 4.68 & 8.95 & 16.68 & 31.74 & {\bf 35.15}\\
 \hline
\end{tabular}

\vspace*{0.5cm}

\begin{tabular}{|l||l|llllll|}
 \hline
 Benchmark & \edt\ version &  1 th. &  2 th. &  4 th. &  8 th. &  16 th. &  32 th.\\
 \hline
 \hline
 & \cncdepend & 2.58 & 4.89 & 8.17 & {\bf 9.36} & 8.38 & 7.36\\
\jacobidoublethreedsevenpointsenlargedoneiteration & \cncsyncget & 3.14 & 4.68 & 7.62 & {\bf 9.67} & 8.71 & 6.53\\
 & \cncunsafeget & 2.77 & 3.55 & 3.26 & 2.90 & 2.67 & 1.78\\
\hline
 & \cncdepend & 2.21 & 4.21 & 7.03 & 11.72 & 17.07 & 19.09\\
\jacobidoublethreedsevenpointsenlarged & \cncsyncget & 2.28 & 4.54 & 7.78 & 14.50 & 25.16 & {\bf 26.37}\\
 & \cncunsafeget & 2.18 & 4.11 & 7.76 & 14.19 & 25.82 & {\bf 27.00}\\
\hline
 & \cncdepend & 2.45 & 4.15 & 4.28 & 5.28 & {\bf 7.47} & {\bf 7.14}\\
\lud & \cncsyncget & 1.05 & 1.94 & 2.45 & 2.41 & 1.76 & 1.59\\
 & \cncunsafeget & 1.38 & 2.47 & 3.34 & 3.09 & 2.71 & 2.50\\
\hline
 & \cncdepend & 4.29 & 8.44 & 15.64 & 25.20 & {\bf 43.30} & 40.63\\
\matmult & \cncsyncget & 4.24 & 8.51 & 15.27 & 28.34 & 42.02 & {\bf 42.17}\\
 & \cncunsafeget & 3.96 & 4.12 & 4.05 & 4.12 & 3.59 & 2.57\\
\hline
 & \cncdepend & 1.36 & 2.75 & 3.68 & 5.61 & 6.72 & {\bf 8.44}\\
\pmatmult & \cncsyncget & 1.26 & 2.62 & 4.55 & 7.25 & 7.62 & 6.24\\
 & \cncunsafeget & 1.28 & 2.48 & 4.66 & 8.37 & {\bf 9.33} & 5.58\\
\hline
& \cncdepend & 0.65 & 1.00 & 1.54 & 2.13 & {\bf 2.57} & {\bf 2.29}\\
\poisson  & \cncsyncget & 0.25 & 0.48 & 0.49 & 0.48 & 0.36 & 0.27\\
 & \cncunsafeget & 0.35 & 0.55 & 0.63 & 0.59 & 0.53 & 0.47\\
\hline
 & \cncdepend & 3.05 & 5.49 & 9.68 & 16.79 & {\bf 22.30} & 19.64\\
\rtmthreed & \cncsyncget & 3.06 & 5.76 & 9.60 & 17.11 & {\bf 21.78} & 18.73\\
 & \cncunsafeget & 3.26 & 2.95 & 3.09 & 3.10 & 2.73 & 2.56\\
%\hline
% & \cncdepend & 0.91 & 1.73 & 3.04 & 5.31 & 9.08 & {\bf 14.30}\\
%\seidel & \cncsyncget & 0.79 & 1.59 & 3.03 & 5.37 & 6.93 & 7.94\\
% & \cncunsafeget & 0.84 & 1.67 & 3.21 & 5.97 & {\bf 10.66} & 8.45\\
\hline
& \cncdepend & 0.57 & 0.98 & 1.43  & 1.93  & {\bf 2.46} & {\bf 2.40} \\
\sor & \cncsyncget & 0.19  & 0.59  & 0.49  & 0.41  & 0.41 & 0.35 \\
& \cncunsafeget & 0.27   & 0.51  & 0.50  & 0.45  & 0.42  & 0.40 \\
\hline
 & \cncdepend & 5.82 & 10.51 & 14.77 & 20.29 & {\bf 22.77} & {\bf 22.26}\\
\strsm & \cncsyncget & 2.75 & 7.78 & 6.00 & 5.01 & 3.84 & 2.94\\
 & \cncunsafeget & 3.50 & 6.23 & 6.82 & 6.39 & 5.69 & 5.07\\
\hline
 & \cncdepend & 1.96 & 3.58 & 4.13 & 5.81 & {\bf 7.90} & {\bf 8.39}\\
\trisolv & \cncsyncget & 1.11 & 2.79 & 2.98 & 2.35 & 1.79 & 1.76\\
 & \cncunsafeget & 1.30 & 2.40 & 3.47 & 3.23 & 1.87 & 2.63\\
%\hline
%\twodimperfectjacobi & \cncdepend & 4.24 & 7.48 & 12.76 & 21.34 & {\bf 29.65} & 23.44\\
% & \cncsyncget & 3.51 & 6.73 & 11.50 & 14.79 & 13.44 & 12.89\\
% & \cncunsafeget & 3.78 & 7.31 & 13.79 & 24.00 & {\bf 25.81} & 15.42\\
\hline
\end{tabular}
%\end{tabular}
\end{center}
}
\caption{\label{table:cnc-dependence} \cnc\ performance in Gflops/s}
\end{table*}

\section{Experiments}
The numbers we present in this experimental section should be viewed as a
baseline performance achievable from a sequential specification automatically
translated into \edts\ before single thread tuning is applied and in the
absence of data and code placement hints to the runtime.
In particular, {\bf no single thread performance optimization} for SIMD, no 
data-layout transformation and no tile size selection heuristic or tuning are
applied except where specified.\footnote{Thus there is a potential 
performance improvement from such transformations beyond the benefits described here.}
The mapping decisions are {\bf the same in all \edt\ cases} except where
specified\footnote{In fact, the generated files differ only in their includes
  and handling of the \Shutdown\ \edt.}. 
Tile sizes for \edts\  in our experiments are fixed to 64 for the innermost
  loops and 16 for non-innermost loops. This is by no means optimal but just a
  heuristic for overdecomposition to occur while keeping a reasonable
  streaming prefetch and single thread performance.
We also compare our results to automatically generated \openmp\ using our
framework which includes a static heuristic for tile size selection~\cite{DBLP:reference/parallel/MeisterVWBLL11,rstream30-Final}. 
The static tile sizes selected for \openmp\ are meant to
load-balance the execution over a statically fixed number of cores and is also
mindful of streaming memory engines.

Table~\ref{table:kernels} gives a characterization of the experiments we
run. For each benchmark, we specify whether it contains symbolic parameters
(and if so how many), the data and iteration space size as well as the number
of \edts\ we generated and the maximum number of floating point operations per 
full \edt(at the tile size granularities described above).
In order to characterize latencies and stress-test the different runtimes, the
experiments we run are diverse in their sizes, running from a mere $53~ms$
in single thread sequential mode
(\jacobidoublethreedsevenpointsenlargedoneiteration) up to $97~s$
(\jacobidoublethreedtwosevenpointsenlarged). 

Experiments are performed on a two socket, eight core per socket Intel
Sandy Bridge E5-2690 @ 2.90GHz running Fedora Core 19. 
Each core is additionally hyperthreaded for a maximum of $32$ threads of execution.
All experiments were run using ``g++-4.8.0 -O3'' and linked with our C++ \ral\
that we targeted to Intel's \cnc\ v0.8, ETI's \swarm\ v0.13, and the
Open Community Runtime (\ocr) v0.8.  
No restriction on processors has been enforced by
using numactl.
We did not observe differences worth reporting with numactl on these examples.

\subsection{\cnc\ Dependence Specification Alternatives}
We present \cnc\ results separately because \cnc\ allows three different modes
of specifying dependences which already give interesting insights. 
The default \ral\ for \cnc\ uses blocking \emph{get} and is referred to as
\cncsyncget. This mechanism may introduce unnecessary overhead.  
We also retargeted the \ral\ to target \cnc's
\emph{unsafe\_get}/\emph{flush\_gets} mechanism to provide more asynchrony.
This mechanism is similar conceptually to the non-blocking gets in
\swarm. 
%The difference comes from the behavior when a get occurs before the
%corresponding put has been posted. The Intel \cnc\  implementation throws a
%C++ exception and rolls back the gets. On the other hand, a \swarm\ task must
%explicitly re-enqueue itself and return. 
A third \cnc\ mechanism is the so-called \emph{depends} mechanism. 
For each task, all its dependences are pre-specified at
the time of task creation. This is similar in philosophy to the \prescriber
\edt\ that we generate automatically for \ocr.
Table~\ref{table:cnc-dependence} shows the baseline performance achieved
by our \cnc\ generated codes when varying the way dependences are specified.
%In the unsafe\_get mode, the \ral\ uses ``prescheduling'' which additionally
%minimizes the number of times a step can be scheduled. At the same time, it
%creates the risk of serialization. This trade-off is reflected in the 
%fully parallel examples \divergencethreedoneiteration,
%\jacobidoublethreedsevenpointsenlargedoneiteration\ and \matmult\ for which
%he unsafe\_get performance is no better than serial.
Unsurprisingly, blocking gets result in significant overheads in cases where
many smaller \edts\ are generated, which require more calls into the runtime.
This effect is not problematic in the larger 3D cases.
More surprising is the fact that \cncdepend\ performs significantly worse in
the cases \gaussseidelthreedsevenpointsenlarged, \gaussseidelthreedtwosevenpointsenlarged, 
\jacobidoublethreedsevenpointsenlarged and \jacobidoublethreedtwosevenpointsenlarged. 
We conjecture this is not due to runtime overhead but to scheduling decisions.
To confirm this, we run the following experiment: we generate 2 levels of
hierarchical \edts\ (which effectively increases the potential runtime
overhead for \cncdepend). In these codes, the non-leaf \worker\ has the
granularity of the two outermost loops, whereas the leaf \worker\ has the
granularity of an original \edt(16-16-16-64).
Despite the increased runtime overhead to manage these nested tasks, we obtain
up to 50\% speedup, as shown in Table~\ref{table:cnc-hierar2}.

\begin{table}
%\hspace{-10mm}
{\scriptsize
\begin{center}
\begin{tabular}{|l||lllll|}
\hline
Benchmark & Type & Data size & Iteration size & \# \edts\ & \# Fp~/~\edt\ \\
\hline
\hline
\divergencethreedoneiteration  & Param. (1) & $256^3$ & $256^3$ & 1~K & 128~K \\
\fdtdtwod & Const. & $1000^2$ & $500 \cdot 1000^2$ & 148~K & 48~K \\
\gaussseideltwodfivepointsenlarged & Param. (2) & $1024^2$ & $256 \cdot 1024^2$ & 16~K & 80~K \\
\gaussseideltwodninepointsenlarged & Param. (2) & $1024^2$ & $256 \cdot 1024^2$ & 16~K & 144~K \\
\gaussseidelthreedtwosevenpointsenlarged & Param. (2) & $256^3$ & $256^4$ & 256~K & 6.75~M \\
\gaussseidelthreedsevenpointsenlarged & Param. (2) & $256^3$ & $256^4$ & 256~K & 1.75~M \\
\jactwod & Const. & $1000^2$ & $1000^3$ & 60~K & 80~K \\
\jacobidoubletwodfivepointsenlarged & Param. (2) & $ 1024^2$ & $256 \cdot 1024^2$ & 16~K & 80~K \\
\jacobidoubletwodninepointsenlarged & Param. (2) & $ 1024^2$ & $256 \cdot 1024^2$ & 16~K & 144~K \\
\jacobidoublethreedtwosevenpointsenlarged & Param. (2) & $256^3$ & $256^4$ & 256~K & 6.75~M \\
\jacobidoublethreedsevenpointsenlargedoneiteration & Param. (2) & $256^3$ & $256^3$ & 1~K & 112~K \\
\jacobidoublethreedsevenpointsenlarged & Param. (2) & $256^3$ & $256^4$ & 256~K & 1.75M \\
\lud & Const. & $1000^2$ & $1000^3/8$ & 60~K & 10~K \\
\matmult & Const. & $1024^2$ & $1024^3$ & 64~K & 32~K \\
\pmatmult & Const. & $256^2$ & $\sum_{i=1}^{256}~i^3$ & 1~K & 32~K \\
\poisson & Const. & $1024^2$ & $32 \cdot 1024^2$ & 11~K & 96~K \\
\rtmthreed & Param. (2) & $256^3$ & $256^3$ & 1~K & 512~K \\
\sor & Const. & $10,000^2$ & $10,000^2$ & 10~M & 5~K \\
\strsm & Const. & $1500^2$ & $1500^3$ & 200~K & 16~K \\
\trisolv &Const. & $1000^2$ & $1000^3$ & 60~K & 16~K \\
\hline
\end{tabular}
\end{center}
}
\caption{\label{table:kernels} Benchmark characteristics}
\end{table}

\subsection{\swarm, \ocr\ and \omp}
We now turn to numbers we obtain with \swarm, \ocr\ and \omp.
We break down the discussion in different categories of benchmarks. The
discussion here is also valid for \cnc. 

\par\noindent
\emph{1) Embarrassingly Parallel Examples} are ones for which no 
runtime dependences are required (\divergencethreedoneiteration,
\jacobidoublethreedsevenpointsenlargedoneiteration\, \rtmthreed\ and \matmult).
The execution times for the first $3$ examples are very low ($53-210$ ms on 1 thread),
and can be viewed as a 
test of runtime latency overhead on very short runs, without dependences.
\matmult\ is a somewhat larger example.
These examples show \swarm\ has a smaller overhead than \cnc\ and \ocr\ for
running parallel tasks, until reaching the hyperthreading mode where
\swarm\ performance consistently drops.
One could argue whether hyperthreading should be accounted for; since it shows
interesting and different behaviors across runtimes, we chose to report it.

\par\noindent
\emph{2) \edt\ granularity.}
\lud, \poisson\ and \sor\ illustrate quite small examples in which the
statically selected tile sizes are not adequate for \edt\ granularity purposes.
In the case of \poisson\ pipeline startup cost is prohibitively expensive; choosing
tile sizes of ${2-32-128}$ yields around $7$ Gflop/s with \ocr\  on 32 threads, a
6x speedup.
In the case of \sor, the tile sizes yield small tasks of merely $1024$ iterations
corresponding to $5$K instructions; selecting larger tile sizes also improves
performance. 
Overall, the examples in this section show that relatively small tile
sizes that achieve overprovisioning may not be beneficial.
We discuss this further for \sor\ and \lud\ in the next section.  

\par\noindent
\emph{3) \omp\ Efficient Examples.}
\strsm\ and \trisolv\ illustrate $2$ cases that mix both parallel and
permutable loops and for which \omp\ performs
significantly better than any of the \edt\ solutions. In this case, we could
narrow the problem down to tile size selection for reuse.
In the case of \strsm, by selecting square tiles of size
$64-64-64$, we were able to obtain up to $76$ Gflop/s with \ocr. 
Unfortunately the performance would not increase further with hyperthreading.
In addition, forcing the \omp\ tile sizes to $16-16-64$ capped the
performance at $50$ Gflop/s. 
In the case of \trisolv, by selecting a tile of size
$64-64-256$,we were able to obtain up to $26$ Gflop/s with \ocr. 
This further emphasizes the need for a proper tile size selection in \edt-based
runtimes. There is a difficult trade-off between over-decomposition, reuse,
single thread performance, streaming prefetch utilization and problem size
that should be solved in a dynamic and adaptive fashion. To allow this
adaptivity in the future, we will generate parametrized tasks that can
recursively decompose and expose the proper parameters to the runtime.

\par\noindent
\emph{4) 2-D and 3-D Time Tiling.}
The remaining examples show the benefit of \edts.
As expected in those cases, performance for \edt\-based codes scales
significantly better than \omp\ performance, especially as the 
Jacobi examples (explicit relaxation scheme) move twice as much memory
as GaussSeidel examples (implicit relaxation scheme) and do not scale as
well from 16 to 32 threads (hyperthreading). Still in the future they are
expected to scale better with more processors because they can be subjected to
diamond tiling\cite{BandishtiPB12} that presents a parallel start-up front and
significantly more parallelism than time tiling schemes. 
%Our motivating
%example in section~\ref{sec:motivation} is an instance of this diamond tiling
%for which the generated \cnc\ was obtained manually.
At 3-D levels of granularity, we see no significant difference between the 3
\edt-based runtimes we targeted.

%\hspace{-10mm}
\begin{table}
{\scriptsize 
\begin{center}
\begin{tabular}{|l||l|llllll|}
  \hline
  Benchmark & version & 1 th. & 2 th. & 4 th. & 8 th. & 16 th. & 32 th.\\
  \hline
  \hline
  \gaussseidelthreedsevenpointsenlarged &  \cncdepend &  1.61 & 3.28 & 6.21 & 11.46 & 21.20 & 32.88\\
  \gaussseidelthreedtwosevenpointsenlarged &  \cncdepend & 1.82 & 3.62 & 6.94 & 13.03 & 25.02 & 34.83\\
  \jacobidoublethreedsevenpointsenlarged &  \cncdepend & 2.12 & 3.91 & 7.44 & 13.19 & 20.18 & 25.11\\
  \jacobidoublethreedtwosevenpointsenlarged &  \cncdepend &2.38 & 4.81 & 8.79 & 16.04 & 30.78 & 33.31\\
  \hline
\end{tabular}    
\end{center}
}
\caption{\label{table:cnc-hierar2} \cnc, two level hierarchy, performance in  Gflops/s}
\end{table}

\begin{table*}
%\hspace{-10mm}
{\scriptsize
  \begin{center}
%  \begin{tabular}{rl}
  \begin{tabular}{|l||l|llllll|}
    \hline
    Benchmark & \edt\ version & 1 th. & 2 th. & 4 th. & 8 th. & 16 th. & 32 th.\\
    \hline
    \hline
    \divergencethreedoneiteration &   \ocr & 2.44 & 3.89 & 4.84 & 6.64 & 6.43 & 5.63\\
%    & \openmpfixed & 2.31 & 2.33 & 4.20 & 6.57 & 9.63 & 8.57\\
    & \openmp & 3.02 & 5.65 & 7.62 & {\bf 8.86} & {\bf 8.76} & 8.46\\
    & \swarm & 1.91 & 4.00 & 6.38 & 8.06 & 8.28 & 2.96\\
    \hline
    \fdtdtwod & \ocr & 1.20 & 2.31 & 4.34 & 8.13 & {\bf 13.96} & {\bf 17.14}\\
%    & \openmpfixed & 1.46 & 2.04 & 3.14 & 5.15 & 7.50 & 7.18\\
    & \openmp & 0.83 & 0.29 & 0.56 & 0.95 & 1.41 & 2.46\\
    & \swarm & 1.17 & 2.07 & 3.91 & 7.46 & 11.91 & 13.75\\
    \hline
    \gaussseideltwodfivepointsenlarged & \ocr & 0.89 & 1.72 & 3.23 & 5.90 & {\bf 10.38} & {\bf 15.04}\\
%    & \openmpfixed & 1.17 & 1.65 & 2.38 & 3.87 & 6.09 & 8.99\\
    & \openmp & 1.13 & 1.14 & 1.16 & 1.19 & 1.22 & 1.28\\
    & \swarm & 0.88 & 1.65 & 3.11 & 5.74 & 9.73 & 3.11\\
    \hline
    \gaussseideltwodninepointsenlarged & \ocr & 0.98 & 1.90 & 3.61 & 6.67 & {\bf 12.05} & {\bf 18.24}\\
%    & \openmpfixed & 1.19 & 1.58 & 2.17 & 3.55 & 5.77 & 8.32\\
    & \openmp & 1.17 & 1.16 & 1.18 & 1.17 & 1.19 & 1.20\\
    & \swarm & 0.96 & 1.85 & 3.50 & 6.51 & 11.51 & 11.88\\
    \hline
    \gaussseidelthreedsevenpointsenlarged & \ocr & 1.55 & 3.04 & 5.87 & 10.91 & 20.72 & {\bf 34.25}\\
%    & \openmpfixed & 1.69 & 2.03 & 2.60 & 4.30 & 7.85 & 13.17\\
    & \openmp & 1.75 & 2.21 & 3.01 & 4.90 & 7.83 & 11.12\\
    & \swarm & 1.56 & 2.93 & 5.64 & 10.64 & 20.29 & {\bf 32.71}\\
    \hline
    \gaussseidelthreedtwosevenpointsenlarged & \ocr & 1.82 & 3.56 & 6.90 & 12.95 & 24.71 & {\bf 37.53}\\
%    & \openmpfixed & 1.98 & 2.24 & 2.54 & 3.72 & 6.50 & 11.75\\
    & \openmp & 2.06 & 3.16 & 5.51 & 10.16 & 18.86 & 29.26\\
    & \swarm & 1.84 & 3.52 & 6.80 & 12.78 & 24.45 & {\bf 37.19}\\
%    \hline
%    \gradientthreedoneiteration & \ocr  & 1.16  & 2.02  & 2.95  & 3.70  & 3.94  & 2.54\\
%%    & \openmpfixed  & 1.41  & 1.37  & 1.92  & 4.10  & 5.70  & 4.99\\
%    & \openmp  & 1.75  & 1.77  & 2.53  & 4.95  & 5.83  & 4.99\\
%    & \swarm  & 0.95  & 2.04  & 3.34  & 4.29  & 4.77  & 2.10\\
    \hline
    \jactwod & \ocr & 4.05 & 7.57 & 14.34 & 25.66 & {\bf 44.81} & {\bf 42.90}\\
%    & \openmpfixed & 4.64 & 7.28 & 12.50 & 20.76 & 33.77 & 26.69\\
    & \openmp & 4.25 & 5.30 & 7.33 & 12.60 & 19.90 & 18.00\\
    & \swarm & 3.67 & 6.12 & 11.34 & 21.40 & 35.51 & 9.37\\
    \hline
    \jacobidoubletwodfivepointsenlarged & \ocr & 1.71 & 3.22 & 6.11 & 11.08 & {\bf 18.98} & {\bf 21.72}\\
%    & \openmpfixed & 0.94 & 1.37 & 2.07 & 3.43 & 5.62 & 8.20\\
    & \openmp & 0.92 & 0.92 & 0.91 & 1.13 & 1.40 & 2.19\\
    & \swarm & 1.63 & 3.15 & 5.84 & 10.63 & 17.58 & 6.15\\
    \hline
    \jacobidoubletwodninepointsenlarged & \ocr & 1.58 & 3.00 & 5.84 & 10.52 & 18.99 & {\bf 21.54}\\
%    & \openmpfixed & 1.20 & 1.71 & 2.60 & 4.51 & 8.04 & 9.43\\
    & \openmp & 1.09 & 1.14 & 1.20 & 1.31 & 1.67 & 2.64\\
    & \swarm & 1.50 & 3.00 & 5.58 & 10.27 & 18.53 & {\bf 19.48}\\
    \hline
    \jacobidoublethreedtwosevenpointsenlarged & \ocr & 2.41 & 4.70 & 8.94 & 16.72 & 31.66 & {\bf 34.48}\\
%    & \openmpfixed & 2.11 & 2.51 & 3.17 & 5.11 & 9.37 & 16.90\\
    & \openmp & 2.43 & 3.43 & 5.66 & 10.36 & 18.87 & 25.95\\
    & \swarm & 2.40 & 4.53 & 8.75 & 16.21 & 30.67 & {\bf 34.51}\\
    \hline
  \end{tabular}

\vspace*{0.5cm}

  \begin{tabular}{|l||l|llllll|}
    \hline
    Benchmark & \edt\ version & 1 th. & 2 th. & 4 th. & 8 th. & 16 th. & 32 th.\\
    \hline
    \hline
    \jacobidoublethreedsevenpointsenlarged & \ocr & 2.18 & 4.14 & 7.80 & 14.17 & 25.50 & {\bf 26.75}\\
%    & \openmpfixed & 1.85 & 2.12 & 2.64 & 4.27 & 7.72 & 12.79\\
    & \openmp & 1.93 & 2.15 & 2.62 & 4.12 & 7.42 & 12.66\\
    & \swarm & 2.12 & 3.81 & 7.32 & 13.74 & 24.84 & {\bf 26.09}\\
    \hline
    \jacobidoublethreedsevenpointsenlargedoneiteration & \ocr & 2.97 & 4.71 & 8.38 & 9.46 & 8.35 & 6.71\\
%    & \openmpfixed & 2.21 & 2.26 & 4.02 & 6.71 & 13.19 & 11.13\\
    & \openmp & 3.33 & 5.70 & 11.61 & {\bf 19.59} & {\bf 17.53} & 13.67\\
    & \swarm & 2.16 & 5.91 & 7.93 & 11.14 & 12.14 & 3.18\\
    \hline
    \lud & \ocr & 1.66 & 2.72 & 5.21 & 7.33 & {\bf 7.67} & 4.91\\
%    & \openmpfixed & 1.46 & 2.14 & 3.29 & 5.47 & 9.23 & 6.22\\
    & \openmp & 0.57 & 0.78 & 0.94 & 0.67 & 0.59 & 0.98\\
    & \swarm & 2.02 & 2.93 & 4.70 & 6.91 & {\bf 7.71} & 1.35\\
    \hline
    \matmult & \ocr & 4.37 & 8.35 & 15.05 & 26.80 & {\bf 45.72} & 43.77\\
%    & \openmpfixed & 1.26 & 2.42 & 4.80 & 9.04 & 17.46 & 14.36\\
    & \openmp & 1.21 & 2.38 & 4.49 & 8.37 & 15.78 & 14.41\\
    & \swarm & 4.46 & 8.58 & 15.49 & 28.87 & {\bf 49.44} & 35.53\\
    \hline
    \pmatmult & \ocr & 1.37 & 2.59 & 4.89 & 8.81 & 14.46 & 15.60\\
%    & \openmpfixed & 1.53 & 1.48 & 2.36 & 4.40 & 8.28 & 14.52\\
    & \openmp & 1.90 & 2.97 & 5.48 & 9.12 & {\bf 15.98} & {\bf 20.14}\\
    & \swarm & 1.35 & 2.66 & 5.05 & 9.35 & 13.55 & 3.82\\
    \hline
    \poisson & \ocr & 0.46 & 0.64 & 1.14 & {\bf 1.71} & 1.43 & 1.00\\
%    & \openmpfixed & 0.84 & 1.16 & 1.77 & 3.15 & 5.53 & 5.97\\
    & \openmp & 1.01 & 0.97 & 1.02 & 0.99 & 0.96 & 0.84\\
    & \swarm & 0.44 & 0.63 & 0.99 & 1.41 & {\bf 1.57} & 0.27\\
    \hline
    \rtmthreed & \ocr & 3.00 & 5.38 & 9.65 & 15.84 & 24.42 & 17.48\\
%    & \openmpfixed & 2.32 & 2.26 & 4.35 & 8.37 & 15.41 & 21.48\\
    & \openmp & 2.40 & 4.59 & 8.03 & 15.77 & {\bf 29.06} & 22.67\\
    & \swarm & 2.83 & 5.95 & 9.76 & 18.02 & {\bf 26.23} & 12.34\\
    \hline
    \sor & \ocr & 0.28  & 0.56  & 0.98  & 1.65  & 1.27  & 0.93 	\\
    %    & \openmpfixed & 1.09 & 1.64 & 2.65 & 4.67 & 8.27 & 14.60\\
    & \openmp & 0.62 & 1.01 & 1.59 & 2.66 & {\bf 4.42} & {\bf 6.62}\\
    & \swarm & 0.26   & 0.45 & 0.68 & 1.17  & 0.86  & 0.22\\
%    \seidel & \ocr & 0.87 & 1.68 & 3.24 & 6.06 & 11.26 & 16.76\\
%%    & \openmpfixed & 1.09 & 1.64 & 2.65 & 4.67 & 8.27 & 14.60\\
%    & \openmp & 1.06 & 1.57 & 2.56 & 4.68 & 8.66 & 15.33\\
%    & \swarm & 0.86 & 1.64 & 3.10 & 5.84 & 10.69 & 4.90\\
    \hline
    \strsm & \ocr & 4.49 & 7.58 & 11.76 & 17.62 & 15.95 & 11.72\\
%    & \openmpfixed & 3.74 & 6.98 & 13.94 & 26.27 & 50.65 & 41.92\\
    & \openmp & 3.66 & 5.60 & 10.52 & 19.84 & {\bf 37.97} & {\bf 39.15}\\
    & \swarm & 2.82 & 4.15 & 7.39 & 13.04 & 17.88 & 2.79\\
    \hline
    \trisolv & \ocr & 1.64 & 2.95 & 4.89 & 7.63 & 7.55 & 5.29\\
%    & \openmpfixed & 2.19 & 4.12 & 7.23 & 15.34 & 27.35 & 23.03\\
    & \openmp & 2.09 & 4.29 & 7.77 & 15.15 & {\bf 28.67} & {\bf 23.28}\\
    & \swarm & 1.56 & 2.84 & 4.88 & 7.88 & 9.77 & 1.37\\
    \hline
  \end{tabular}    
%  \end{tabular}    
  \end{center}
\caption{\label{table:cnc-hierar} \swarm, \ocr\ and \omp\ performance in Gflops/s}
}
\end{table*}

\subsection{Effects of \edt\ Granularity}
We further investigated the examples of \lud\ and \sor\ on which
\edt\ performance was lower than we expected. We investigated a few different tile sizes 
as well as 2 levels of granularity (for \lud). The granularity parameter 
represents the number of loops in an \edt.  Figure~\ref{table:ocr-size} shows
there is a fine trade-off between \edt\ granularity, number of \edts\ and cost
of managing these \edts. 
\begin{table}
%\hspace{-10mm}
{\scriptsize
  \begin{center}
  \begin{tabular}{|l||l|l|llllll|}
    \hline
    Benchmark & Sizes & Gran. & 1 th. & 2 th. & 4 th. & 8 th. & 16 th. & 32 th.\\
    \hline
    \hline
\lud & 16-16-16 & 3 &  0.92 & 1.54 & 2.55 & 3.49 & 2.04 & 1.41\\
\lud & 16-16-16 & 4 &  1.85 & 3.66 & 6.63 & 11.44 & {\bf 16.68} & 9.56\\
\lud & 64-64-64 & 3 &  2.41 & 4.66 & 8.01 & 13.34 & 14.15 & 10.80\\
\lud & 64-64-64 & 4 &  1.94 & 3.63 & 5.73 & 6.54 & 6.22 & 3.46\\
\lud & 10-10-100 & 3 & 1.90 & 3.57 & 6.55 & 11.64 & {\bf 16.25} & 10.67\\
\lud & 10-10-100 & 4 & 1.83 & 3.56 & 6.63 & 11.64 & 14.39 & 9.93\\
\hline
\hline
%\sor & 32-1024 &  2 & 0.66 & 1.24 & 2.00 & 3.64 & 4.74 & 4.02\\
\sor & 100-100 &  2 & 0.64 & 1.21 & 2.28 & 4.15 & 6.85 & 5.18\\
\sor & 100-1000 & 2 & 0.64 & 1.28 & 2.09 & 3.46 & 4.84 & 4.18\\
%\sor & 128-1024 & 2 & 0.67 & 1.26 & 2.20 & 3.59 & 4.72 & 3.46\\
\sor & 200-200 &  2 & 0.63 & 1.26 & 2.31 & 4.28 & {\bf 7.04} & {\bf 8.16}\\
\sor & 1000-1000 &2 &0.67 & 1.21 & 2.03 & 2.79 & 2.97 & 2.73\\
    \hline
  \end{tabular}    
  \end{center}
\caption{\label{table:ocr-size} \ocr, tile size exploration, performance in Gflops/s}
}
\end{table}    
To confirm the runtime overhead as \edts\ shrink in size, we additionally
collected performance hotspots using Intel Vtune amplxe-cl for 
\lud 16-16-16 with granularity 3 and 4 at 16 threads. First, we did not
observe templated expressions calculations appearing, confirming the low
extra overhead of our solution.
Second, in the case of granularity 4, more than 85\% of the non-idle time is spent
executing work, the rest being spent mostly in the \ocr\ dequeInit function.
However, at the finer granularity, the ratio of effective work drops to merely
10\% stealing and queue management taking up to 80\%.
% with deque\_steal, hcSchedulerTake, workpileIteratorNext,
%workpileIteratorHasNext and ptrReleaseGuid taking more than 80\% of the time.
The drop in performance between 16-16-16 and 10-10-100 suggests there is a
critical threshold, possibly linked to last-level cache sizes, at which the overhead of
\ocr\ increases substantially.
While these initial performance results are encouraging when the granularity
selection is adequate, the overhead introduced to manage fine-grained tasks is too
high at the moment to allow the envisioned level of overdecomposition.
Still the \ocr\ implementation is in the early stages and increasing
performance is to be expected.

\section{Related Work}
The most closely related work to ours is~\cite{Baskaran2009}, but our contribution presents significant differences with this work. 

First, our scheme is generic, targets three different runtimes and is easily
extensible to other runtimes based on task graphs like \ocr\ or on tag-tuples
like \cnc\ and \swarm.
The experiments section shows the variability between runtimes and the
benefit of a nimble strategy.
Second, our scheme is decentralized and fully asynchronous in the creation of
tasks and the dependences between them. 
Baskaran's solution must first construct the full graph and then only begin
useful work. Considering Amdahl's law, our solution is expected to scale
better on larger numbers of processors and distributed memory.
Third, our baseline dependence specification mechanism is scalable at both
compile-time and runtime by virtue of exploiting loop types and dependence
information on restructured loops available from the scheduler. 
%It also supports \edt\ graphs coming from
%imperfectly nested loops thanks to our synchronization using Startup and
%Shutdown \edts.
%Baskaran's mechanism is subject to severe
%scalability constraints because the exact computation of last-write
%information on general tiled loop nests is very expensive. Even more so in the
%context of imperfectly nested loops~\cite{feautrier91dataflow}.
Last, loop type information is a general compiler concept and has a chance to
be extended to more general dependence analyses~\cite{Oancea2012}. 
Our expression template dependence resolution and hierarchical decomposition
mechanism are first steps in this direction.  

There are a number of other runtimes that our mapping algorithms could
target.  Most successfully, the QUARK
runtime~\cite{DBLP:journals/concurrency/HaidarLYD11,quark11}, speeds
the PLASMA linear algebra
library~\cite{DBLP:reference/parallel/DongarraL11f} with dynamic task
scheduling and a task-oriented execution model.  However, PLASMA is
parallelized to QUARK {\em by hand}.  Our mapping approach may be used
to more rapidly and portably {\em automatically} generate
task-oriented implementations of linear algebra to QUARK.  Then, it
could be used to regenerate implementations of such a linear algebra
library taking advantage of the features of CnC, SWARM, and OCR.
Furthermore, our approach is directly oriented toward porting the
library for impending architectural changes from exascale, such as very deep memory hierarchies.
Other EDT oriented runtimes that our mapping approach could
target include The Qthreads Library~\cite{DBLP:conf/ipps/WheelerMT08} or
HPX~\cite{DBLP:journals/sigmetrics/TabbalABKS11}.

\section{Conclusions}
We have demonstrated the first fully automatic solution for generating 
event-driven, tuple-space based programs from a sequential C 
specification for multiple \edt-based runtimes.
 Our solution performs hierarchical mapping and exploits
hierarchical async-finishes. Our solution uses 
auto-parallelizing compiler technology to target three different runtimes
relying on event-driven tasks (\edts) via a runtime-agnostic C++ layer, which
we have retargeted to Intel's Concurrent Collections (\cnc), ETI's \swarm,
and the Open Community Runtime (\ocr). 
We have further demonstrated the performance improvements achievable with these \edt-based
runtimes and pinpointed some current weaknesses related to fine-grained
\edt\ management overhead.
Our solution takes advantage of parallel and permutable loops to abstract
aggregate dependences between \edts. The notion of permutability is a general
compiler concept and the transformations we propose can be extended to more
general programs than we currently handle. Lastly, our templated expression
based multi-dimensional spaces can be extended to decompose recursively and
adapt at runtime, which will be the subject of future work.

%\appendix
%\section{Appendix Title}
%
%This is the text of the appendix, if you need one.
%
%\acks
%
%Acknowledgments, if needed.

% We recommend abbrvnat bibliography style.
% \bibliographystyle{abbrvnat}
\bibliographystyle{alpha}

% The bibliography should be embedded for final submission.

\bibliography{pca}

\newcommand{\etalchar}[1]{$^{#1}$}
\begin{thebibliography}{MVW{\etalchar{+}}11}

\bibitem[Bas04a]{Bas04b}
C.~Bastoul.
\newblock Code generation in the polyhedral model is easier than you think.
\newblock In {\em PACT'04}, pages 7--16, Juan-les-Pins, September 2004.

\bibitem[Bas04b]{DBLP:conf/IEEEpact/Bastoul04}
C{\'e}dric Bastoul.
\newblock Code generation in the polyhedral model is easier than you think.
\newblock In {\em IEEE PACT}, pages 7--16. IEEE Computer Society, 2004.

\bibitem[BBC{\etalchar{+}}10]{CnCZoran}
Zoran Budimlic, Michael~G. Burke, Vincent Cav{\'e}, Kathleen Knobe, Geoff
  Lowney, Ryan Newton, Jens Palsberg, David~M. Peixotto, Vivek Sarkar, Frank
  Schlimbach, and Sagnak Tasirlar.
\newblock Concurrent collections.
\newblock {\em Scientific Programming}, 18(3-4):203--217, 2010.

\bibitem[BHRS08]{bondhugula08}
U.~Bondhugula, A.~Hartono, J.~Ramanujan, and P.~Sadayappan.
\newblock A practical automatic polyhedral parallelizer and locality optimizer.
\newblock In {\em ACM SIGPLAN Programming Languages Design and Implementation
  (PLDI '08)}, Tucson, Arizona, June 2008.

\bibitem[BHT{\etalchar{+}}10]{DBLP:conf/cgo/BaskaranHTHRS10}
Muthu~Manikandan Baskaran, Albert Hartono, Sanket Tavarageri, Thomas Henretty,
  J.~Ramanujam, and P.~Sadayappan.
\newblock Parameterized tiling revisited.
\newblock In Andreas Moshovos, J.~Gregory Steffan, Kim~M. Hazelwood, and
  David~R. Kaeli, editors, {\em CGO}, pages 200--209. ACM, 2010.

\bibitem[BJK{\etalchar{+}}96]{Blumofe96}
Robert~D. Blumofe, Christopher~F. Joerg, Bradley~C. Kuszmaul, Charles~E.
  Leiserson, Keith~H. Randall, and Yuli Zhou.
\newblock Cilk: An efficient multithreaded runtime system.
\newblock {\em J. Parallel Distrib. Comput.}, 37(1):55--69, 1996.

\bibitem[Boa08]{openmp}
OpenMP Architecture~Review Board.
\newblock The openmp specification for parallel programming {\tt
  http://www.openmp.org}, 2008.

\bibitem[BPB12]{BandishtiPB12}
Vinayaka Bandishti, Irshad Pananilath, and Uday Bondhugula.
\newblock Tiling stencil computations to maximize parallelism.
\newblock In {\em SC}, page~40, 2012.

\bibitem[BVB{\etalchar{+}}09]{Baskaran2009}
Muthu~Manikandan Baskaran, Nagavijayalakshmi Vydyanathan, Uday Bondhugula,
  J.~Ramanujam, Atanas Rountev, and P.~Sadayappan.
\newblock Compiler-assisted dynamic scheduling for effective parallelization of
  loop nests on multicore processors.
\newblock In Daniel~A. Reed and Vivek Sarkar, editors, {\em PPOPP}, pages
  219--228. ACM, 2009.

\bibitem[CA88]{DBLP:conf/isca/CullerA88}
David~E. Culler and Arvind.
\newblock Resource requirements of dataflow programs.
\newblock In Howard~Jay Siegel, editor, {\em ISCA}, pages 141--150. IEEE
  Computer Society, 1988.

\bibitem[Car13]{Carter2013Short}
Nicholas Carter.
\newblock et al. runnemede: An architecture for ubiquitous high-performance
  computing.
\newblock In {\em Proceedings of the 2013 IEEE 19th International Symposium on
  High Performance Computer Architecture (HPCA)}, HPCA '13, pages 198--209,
  Washington, DC, USA, 2013. IEEE Computer Society.

\bibitem[CBF95]{collard95fuzzy}
J.-F. Collard, D.~Barthou, and P.~Feautrier.
\newblock Fuzzy array dataflow analysis.
\newblock In {\em Proceedings of the 5th {ACM} {SIGPLAN} Symposium on
  Principles and Practice of Parallel Programming}, pages 92--101, Santa
  Barbara, California, July 1995.

\bibitem[CI95]{Beatrice95interprocedural}
B\'eatrice Creusillet and Francois Irigoin.
\newblock {I}nterprocedural {A}rray {R}egion {A}nalyses.
\newblock In {\em Eighth International Workshop on Languages and Compilers for
  Parallel Computing ({LCPC}'95)}, pages 4--1 to 4--15, August 1995.

\bibitem[Den74]{DBLP:conf/programm/Dennis74}
Jack~B. Dennis.
\newblock First version of a data flow procedure language.
\newblock In Bernard Robinet, editor, {\em Symposium on Programming}, volume~19
  of {\em Lecture Notes in Computer Science}, pages 362--376. Springer, 1974.

\bibitem[DL11]{DBLP:reference/parallel/DongarraL11f}
Jack Dongarra and Piotr Luszczek.
\newblock Plasma.
\newblock In Padua \cite{DBLP:reference/parallel/2011}, pages 1568--1570.

\bibitem[Fea88]{feautrier88parametric}
P.~Feautrier.
\newblock Parametric integer programming.
\newblock {\em RAIRO-Recherche Op\'erationnelle}, 22(3):243--268, 1988.

\bibitem[Fea91]{feautrier91dataflow}
P.~Feautrier.
\newblock Dataflow analysis of array and scalar references.
\newblock {\em International Journal of Parallel Programming}, 20(1):23--52,
  February 1991.

\bibitem[Fea92]{feautrier92some}
P.~Feautrier.
\newblock Some efficient solutions to the affine scheduling problem. {P}art
  {I}. {O}ne-dimensional time.
\newblock {\em International Journal of Parallel Programming}, 21(5):313--348,
  October 1992.

\bibitem[FLPR99]{DBLP:conf/focs/FrigoLPR99}
Matteo Frigo, Charles~E. Leiserson, Harald Prokop, and Sridhar Ramachandran.
\newblock Cache-oblivious algorithms.
\newblock In {\em FOCS}, pages 285--298. IEEE Computer Society, 1999.

\bibitem[GFL99]{Griebl99indexset}
M.~Griebl, P.~Feautrier, and C.~Lengauer.
\newblock Index set splitting.
\newblock {\em International Journal of Parallel Programming}, 28:607--631,
  July 1999.

\bibitem[Gri04]{griebl04habilitation}
Martin Griebl.
\newblock Automatic parallelization of loop programs for distributed memory
  architectures, 2004.
\newblock Habilitation thesis.

\bibitem[GVB{\etalchar{+}}06]{GV06}
S.~Girbal, N.~Vasilache, C.~Bastoul, A.~Cohen, D.~Parello, M.~Sigler, and
  O.~Temam.
\newblock Semi-automatic composition of loop transformations for deep
  parallelism and memory hierarchies.
\newblock {\em Int. J. Parallel Program.}, 34(3):261--317, June 2006.

\bibitem[HBB{\etalchar{+}}09]{HBBCKNRS09}
A.~Hartono, M.~Baskaran, C.~Bastoul, A.~Cohen, S.~Krishnamoorthy, B.~Norris,
  J.~Ramanujam, and P.~Sadayappan.
\newblock Parametric multi-level tiling of imperfectly nested loops.
\newblock In {\em Proceedings of the ACM International Conference on
  Supercomputing ({ICS'09})}, pages 147--157, Yorktown Heights, New York, June
  2009.

\bibitem[HLYD11]{DBLP:journals/concurrency/HaidarLYD11}
Azzam Haidar, Hatem Ltaief, Asim YarKhan, and Jack Dongarra.
\newblock Analysis of dynamically scheduled tile algorithms for dense linear
  algebra on multicore architectures.
\newblock {\em Concurrency and Computation: Practice and Experience},
  24(3):305--321, 2011.

\bibitem[IHBZ10]{DBLP:conf/ipps/IancuHBZ10}
Costin Iancu, Steven~A. Hofmeyr, Filip Blagojevic, and Yili Zheng.
\newblock Oversubscription on multicore processors.
\newblock In {\em IPDPS}, pages 1--11. IEEE, 2010.

\bibitem[Inta]{CnC}
Intel.
\newblock Concurrent collections.
\newblock
  http://software.intel.com/en-us/articles/intel-concurrent-collections-for-cc/.

\bibitem[Intb]{swarm}
ET~International.
\newblock {SW}ift adaptive runtime machine.
\newblock http://www.etinternational.com/index.php/products/swarmbeta/.

\bibitem[IT88]{irigoin88}
F.~Irigoin and R.~Triolet.
\newblock Supernode partitioning.
\newblock In {\em Proceedings of the 15th ACM SIGPLAN-SIGACT Symposium on
  Principles of programming languages}, pages 319--329, New York, NY, USA,
  January 1988. ACM Press.

\bibitem[JCP{\etalchar{+}}12]{JimboreanCPML12}
Alexandra Jimborean, Philippe Clauss, Beno\^{\i}t Pradelle, Luis Mastrangelo,
  and Vincent Loechner.
\newblock Adapting the polyhedral model as a framework for efficient
  speculative parallelization.
\newblock In {\em PPOPP}, pages 295--296, 2012.

\bibitem[KAH{\etalchar{+}}12]{DBLP:conf/dac/KaulAHAKB12}
Himanshu Kaul, Mark Anders, Steven Hsu, Amit Agarwal, Ram Krishnamurthy, and
  Shekhar Borkar.
\newblock Near-threshold voltage (ntv) design: opportunities and challenges.
\newblock In Patrick Groeneveld, Donatella Sciuto, and Soha Hassoun, editors,
  {\em DAC}, pages 1153--1158. ACM, 2012.

\bibitem[KBB{\etalchar{+}}07]{Sriram2007}
Sriram Krishnamoorthy, Muthu~Manikandan Baskaran, Uday Bondhugula,
  J.~Ramanujam, Atanas Rountev, and P.~Sadayappan.
\newblock Effective automatic parallelization of stencil computations.
\newblock In {\em PLDI}, pages 235--244, 2007.

\bibitem[KBC{\etalchar{+}}08]{Kogge:Exascale08}
P.~M. Kogge, Shekhar Borkar, William~W. Carlson, William~J. Dally, Monty
  Denneau, Paul~D. Franzon, Stephen~W. Keckler, Dean Klein, Robert~F. Lucas,
  Steve Scott, Allan~E. Snavely, Thomas~L. Sterling, R.~Stanley Williams,
  Katherine~A. Yelick, William Harrod, Daniel~P. Campbell, Kerry~L. Hill,
  Jon~C. Hiller, Sherman Karp, Mark~A. R.s, and Alfred~J. Scarpelli.
\newblock {Exascale Study Group: Technology Challenges in Achieving Exascale
  Systems}.
\newblock Technical report, DARPA, 2008.

\bibitem[LLM{\etalchar{+}}08]{rstream30-Final}
R.~Lethin, A.~Leung, B.~Meister, P.~Szilagyi, N.~Vasilache, and D.~Wohlford.
\newblock Final report on the {R}-{S}tream 3.0 compiler {DARPA/AFRL Contract \#
  F03602-03-C-0033, DTIC AFRL-RI-RS-TR-2008-160}.
\newblock Technical report, Reservoir Labs, Inc., May 2008.

\bibitem[LRW91]{DBLP:conf/asplos/LamRW91}
Monica~S. Lam, Edward~E. Rothberg, and Michael~E. Wolf.
\newblock The cache performance and optimizations of blocked algorithms.
\newblock In David~A. Patterson, editor, {\em ASPLOS}, pages 63--74. ACM Press,
  1991.

\bibitem[LVML09]{pat.locality}
Allen~K. Leung, Nicolas~T. Vasilache, Benoit Meister, and Richard~A. Lethin.
\newblock Methods and apparatus for joint parallelism and locality optimization
  in source code compilation, September 2009.

\bibitem[MS68]{DBLP:journals/cacm/MyerS68}
T.~H. Myer and Ivan~E. Sutherland.
\newblock On the design of display processors.
\newblock {\em Commun. ACM}, 11(6):410--414, 1968.

\bibitem[MVW{\etalchar{+}}11]{DBLP:reference/parallel/MeisterVWBLL11}
Beno\^{\i}t Meister, Nicolas Vasilache, David Wohlford, Muthu~Manikandan
  Baskaran, Allen Leung, and Richard Lethin.
\newblock R-stream compiler.
\newblock In Padua \cite{DBLP:reference/parallel/2011}, pages 1756--1765.

\bibitem[OR12]{Oancea2012}
Cosmin~E. Oancea and Lawrence Rauchwerger.
\newblock Logical inference techniques for loop parallelization.
\newblock In {\em Proceedings of the 33rd ACM SIGPLAN conference on Programming
  Language Design and Implementation}, PLDI '12, pages 509--520, New York, NY,
  USA, 2012. ACM.

\bibitem[Pad11]{DBLP:reference/parallel/2011}
David~A. Padua, editor.
\newblock {\em Encyclopedia of Parallel Computing}.
\newblock Springer, 2011.

\bibitem[PSP98]{MPI-user-guide}
Department of~Mathematics P.~S.~Pacheco.
\newblock {\em A User's Guide to {MPI}}, March 1998.

\bibitem[PW92]{Pugh:1992}
William Pugh and David Wonnacott.
\newblock Eliminating false data dependences using the omega test.
\newblock In {\em Proceedings of the ACM SIGPLAN 1992 conference on Programming
  language design and implementation}, PLDI '92, pages 140--151, New York, NY,
  USA, 1992. ACM.

\bibitem[RAO98]{RauchwergerAO98}
Lawrence Rauchwerger, Francisco Arzu, and Koji Ouchi.
\newblock Standard templates adaptive parallel library (stapl).
\newblock In {\em LCR}, pages 402--409, 1998.

\bibitem[Rei07]{TBB}
James Reinders.
\newblock {\em Intel threading building blocks - outfitting C++ for multi-core
  processor parallelism}.
\newblock O'Reilly, 2007.

\bibitem[RKRS07]{RKRSPLDI07}
Lakshminarayanan Renganarayanan, DaeGon Kim, Sanjay~V. Rajopadhye, and
  Michelle~Mills Strout.
\newblock Parameterized tiled loops for free.
\newblock In {\em PLDI}, pages 405--414, 2007.

\bibitem[SGS{\etalchar{+}}93]{DBLP:conf/isca/SpertusGSECD93}
Ellen Spertus, Seth~Copen Goldstein, Klaus~E. Schauser, Thorsten von Eicken,
  David~E. Culler, and William~J. Dally.
\newblock Evaluation of mechanisms for fine-grained parallel programs in the
  j-machine and the cm-5.
\newblock In Alan~Jay Smith, editor, {\em ISCA}, pages 302--313. ACM, 1993.

\bibitem[TAB{\etalchar{+}}11]{DBLP:journals/sigmetrics/TabbalABKS11}
Alexandre Tabbal, Matthew Anderson, Maciej Brodowicz, Hartmut Kaiser, and
  Thomas~L. Sterling.
\newblock Preliminary design examination of the parallex system from a software
  and hardware perspective.
\newblock {\em SIGMETRICS Performance Evaluation Review}, 38(4):81--87, 2011.

\bibitem[TCK{\etalchar{+}}11]{Tang:2011}
Yuan Tang, Rezaul~Alam Chowdhury, Bradley~C. Kuszmaul, Chi-Keung Luk, and
  Charles~E. Leiserson.
\newblock The {Pochoir} {Stencil} {Compiler}.
\newblock In {\em SPAA '11: The 23rd ACM Symposium on Parallelism in Algorithms
  and Architectures}, San Jose, California, June 2011.

\bibitem[tea]{OCR}
The~OCR team.
\newblock The ocr project.
\newblock https://01.org/projects/open-community-runtime.

\bibitem[WMT08]{DBLP:conf/ipps/WheelerMT08}
Kyle~B. Wheeler, Richard~C. Murphy, and Douglas Thain.
\newblock Qthreads: An {API} for programming with millions of lightweight
  threads.
\newblock In {\em IPDPS}, pages 1--8. IEEE, 2008.

\bibitem[YAK00]{DBLP:conf/pldi/YiAK00}
Qing Yi, Vikram~S. Adve, and Ken Kennedy.
\newblock Transforming loops to recursion for multi-level memory hierarchies.
\newblock In Monica~S. Lam, editor, {\em PLDI}, pages 169--181. ACM, 2000.

\bibitem[YKD11]{quark11}
A.~YarKhan, J.~Kurzak, and J.~Dongarra.
\newblock Quark users' guide: Queueing and runtime for kernels.
\newblock Technical Report ICL-UT-11-02, University of Tennessee Innovative
  Computing Laboratory, 2011.

\end{thebibliography}

\end{document}